\documentclass[onecolumn,epsf,prd,floatfix,nofootinbib,showpacs]{revtex4}
\usepackage{graphicx}
\usepackage{epsfig}
\usepackage{dcolumn}
\usepackage{bm}
\usepackage{latexsym}
\newcommand{\dif}{\mathrm{d}}

\begin{document}

\title{Molecular Bremsstrahlung Radiation at GHz Frequencies in Air}

\author{Imen Al Samarai$^1$, Corinne B\'erat$^2$, Olivier Deligny$^3$,  Antoine Letessier-Selvon$^1$, Fran\c{c}ois Montanet$^2$, Mariangela Settimo$^1$, Patrick Stassi$ ^2$}
\affiliation{
$^1$ Laboratoire de Physique Nucl\'eaire et de Hautes \'Energies, CNRS-IN2P3 \& Universit\'es Pierre et Marie Curie \& Paris Diderot, Paris, France\\
$^2$ Laboratoire de Physique Subatomique et de Cosmologie, Universit\'e Grenoble-Alpes, CNRS/IN2P3, 38026 Grenoble Cedex, France\\
$^3$ Institut de Physique Nucl\'{e}aire, CNRS-IN2P3, Univ. Paris-Sud, Universit\'{e} Paris-Saclay, 91406 Orsay Cedex, France
}

\begin{abstract}
A detection technique for ultra-high energy cosmic rays, complementary to the fluorescence technique, would be the use of the molecular Bremsstrahlung radiation emitted by low-energy ionization electrons left after the passage of the showers in the atmosphere. In this article, a detailed estimate of the spectral intensity of photons at ground level originating from this radiation is presented. The spectral intensity expected from the passage of the high-energy electrons of the cascade is also estimated. The absorption of the photons in the plasma of electrons/neutral molecules is shown to be negligible. The obtained spectral intensity is shown to be $2\times10^{-21}~$W~cm$^{-2}$~GHz$^{-1}$ at 10~km from the shower core for a vertical shower induced by a proton of $10^{17.5}$~eV. In addition, a recent measurement of Bremsstrahlung radiation in air at gigahertz frequencies from a beam of electrons produced at 95~keV by an electron gun is also discussed and reasonably reproduced by the model.
\end{abstract}


\maketitle


\section{Introduction}

The origin and nature of ultra-high energy cosmic rays still remain to be elucidated despite the recent progress provided by the data collected at the Pierre Auger Observatory and the Telescope Array~\cite{Kampert2014}. This is due, in particular, to the extremely low flux of particles at these energies. Currently, the most direct way to infer the nature of ultra-high energy cosmic rays relies on the observation of the longitudinal profile of the atmospheric cosmic-ray showers to measure its maximum of development. This is mainly achieved by making use of telescopes detecting the fluorescence light emitted by nitrogen molecules excited by low-energy ionization electrons left after the passage of the electromagnetic cascade in the atmosphere. This technique, however, can only be used on moonless nights. This results in a $\simeq 10\%$ duty cycle. Together with the low flux of particles, this makes the study of the cosmic ray composition above a few $10^{19}$~eV very challenging.

The search for new techniques able to provide measurements of the electromagnetic content of the cascade with quality comparable to the fluorescence detectors but with 100\% duty cycle is thus an intense field of research and development.  Molecular Bremsstrahlung radiation in the gigahertz frequency range from low-energy ionization electrons left after the passage of the shower provides an interesting mechanism to measure the longitudinal profile of the showers. This is because the emission of gigahertz photons from electrons with energies of the order of one electron volt is isotropic. 

To this aim, measurements in laboratory with beams of electrons have been and are still performed~\cite{Gorham2008,MAYBE,AMY,Conti2014} as well as experimental setups to detect extensive air showers in this way~\cite{MIDAS,Gaior2013,CROME}. Laboratory measurements have led to contradictory results, while setups to detect extensive air showers have led to the detection of a few events only. In addition, the interpretation of these detected events in terms of molecular Bremsstrahlung radiation is still challenging. The detected signals reported in~\cite{Gaior2013,CROME} with antennas oriented vertically or nearly vertically have been observed close to the shower core, where other mechanisms apart from molecular Bremsstrahlung radiation could contribute the observed signals~\cite{AlvarezMuniz2011,Werner2012,AlvarezMuniz2012}. Hence, the magnitude of the expected signals at ground level is still an open question, previously addressed in~\cite{AlSamarai2015}. Compared to this previous work, several significant improvements of the model are introduced in this study, such as a comprehensive description of the collision processes for electron energies smaller than 1~eV, a detailed estimate of the electromagnetic effects in the plasma is given based on a perturbative approach, a detailed description of the absorption effects in the plasma of electron/neutral molecules, and an estimate of the spectral intensity\footnote{In this paper, the term `intensity` refers to as a power per surface unit. The `spectral intensity` is thus the intensity per frequency unit.} expected at ground originating from the high-energy electrons of the showers. 

This paper is organized as follows. In section~\ref{production}, we present the general material needed to estimate the photon production rate from Bremsstrahlung radiation, accounting for suppression effects of Bremsstrahlung for high-energy electrons. The molecular Bremsstrahlung radiation in extensive air showers is then detailed in section~\ref{mbr1}, where a comprehensive treatment of the Boltzmann equation governing the time evolution of the velocity distribution of the ionization electrons is presented. Also, the eventual absorption of the emitted photons in the plasma is addressed in detail through the estimation of the dielectric coefficient of the plasma. In parallel, an estimate of the radiation produced by the high-energy electrons of the cascade is presented for the first time. In section ~\ref{mbr2}, we also confront the model with the recent measurement performed with an electron gun in the laboratory~\cite{Conti2014}. In particular, we show that the model can reproduce reasonably well the order of magnitude of the observed power. This requires a different regime of the Boltzmann equation to be considered. This is due to the much larger density of electrons in this kind of experiment than in extensive air showers, which forces us to consider the electromagnetic term in addition to the collision one in the Boltzmann equation. Conclusions are finally given in section~\ref{conclusion}.

\section{Photon Production Rate from Bremsstrahlung}
\label{production}

\subsection{Photon Production Rate from Bremsstrahlung}
\label{rate}

Electrons can produce low-energy photons by Bremsstrahlung through the process of quasi-elastic collisions with neutral molecules in the atmosphere. The production of photons with energies $h\nu$, with $h$ the Planck constant and $\nu$ the frequency, corresponds to transitions between unquantized energy states of the electrons. We anticipate that the production of photons by Bremsstrahlung is a rare process compared to the other relevant processes entering into play for the electrons. In these conditions, the production rate $r_\gamma$ of photons with energy $h\nu$ per volume unit and per frequency unit can be considered proportional to the density $\rho_m\mathcal{N}_A/A$ of nitrogen and oxygen molecules in air. Here, $\rho_m$ is the mass density, $\mathcal{N}_A$ the Avogadro number and $A$ the mass number.  The production rate is also governed by the electron differential flux and by the double differential cross section for the Bremsstrahlung process for an electron under incidence $(\chi^{\prime},\psi^{\prime})$ with energy $T$ to produce a photon in the direction $(\chi,\psi)$ with energy $h\nu$. For a differential flux of electrons $\phi^\Omega_{\mathrm{e}}$, this leads 
to the following expression:
\begin{equation}
\label{eqn:rate}
r_{\gamma}(\chi,\psi,\nu)=\frac{\rho_m\mathcal{N}_A}{A}~\int \dif T\int\dif \Omega^{\prime}~\phi^\Omega_{\mathrm{e}}(T,\chi^{\prime},\psi^{\prime})~\frac{\dif ^2\sigma}{\dif \nu\dif \Omega}\left(T,\nu,\chi,\psi,\chi^{\prime},\psi^{\prime}\right).
\end{equation}
The solid angle integration accounts for the angular dispersion of the electrons emitting in the considered direction $(\chi,\psi)$. This equation, with additional dependencies in time and position in space, will be of central importance throughout the paper to estimate the expected power in any antenna receiver located at a distance $R$ from the emission point. One of the ingredients is the Bremsstrahlung double differential cross section, that is overviewed in details below.

\subsection{Bremsstrahlung Differential Cross Section}
\label{xs}

For $h\nu/T\ll 1$ and $T\leq 100~$keV, the angular dependence of the double differential cross section is uniform. On the flip side, still for $h\nu/T\ll 1$ but $T\geq 100~$keV, this angular dependence is not uniform any longer and can be approximated by the simple following expression derived in the framework of classical field theory~\cite{Jackson}:
\begin{equation}
\label{eqn:dsigma_ff}
\frac{\dif ^2\sigma}{\dif \nu\dif \Omega}(T,\nu,\omega)=\frac{3}{16\pi}\left(\frac{1-\beta^2(T)}{(1-\beta(T)\cos{\omega})^2}\right)\left(1+\left(\frac{\beta(T)-\cos{\omega}}{1-\beta(T)\cos{\omega}}\right)^2\right)~\frac{\dif \sigma}{\dif \nu}\left(T,\nu\right),
\end{equation}
with $\beta(T)$ the relativistic factor, and $\omega$ the space angle between the direction of the electron $(\chi^\prime,\psi^\prime)$ and the one of the emitted photon $(\chi,\psi)$:
\begin{equation}
\label{eqn:omega}
\omega=\arccos{\left(\sin{\chi}\sin{\chi^\prime}\cos{(\psi-\psi^\prime)}+\cos{\chi}\cos{\chi^\prime}\right)}.
\end{equation}
The factor in front of $\dif \sigma/\dif \nu$ is normalized when integrated over solid angle. It implies an anisotropic emission which is  peaked in the forward direction of the electrons. 

Although the Bremsstrahlung process is well understood, screening effects prevent quantitative calculations of the cross section without approximations which depend on the kinetic energy of the electron and on the emitted energy of the photon. For $T$ up to a few tens of electron volts, screening effects are important and the differential cross section $\dif \sigma/\dif \nu$ is well described by the following expression obtained in~\cite{Kasyanov1961}:
\begin{equation}
\label{eqn:xs-freefree}
\frac{\dif \sigma}{\dif \nu}(T,\nu)=\frac{4\alpha^3}{3\pi R_y\nu}\left(1-\frac{h\nu}{2T}\right)\sqrt{1-\frac{h\nu}{T}}~T\sigma_m(T),
\end{equation}
where $\alpha$ is the fine-structure constant, $R_y$ the Rydberg constant and $\sigma_m(T)$ the electron transfer cross section taken from tabulated data in~\cite{JILA}. For kinetic energies $T$ greater than a few hundreds electron volts, the differential cross section $\dif \sigma/\dif \nu$ is generally written in terms of the `scaled energy loss cross section` $\chi(Z,T,h\nu/T)$:
\begin{equation}
\label{eqn:xs-scaled}
\frac{\dif \sigma}{\dif \nu}(T,\nu)=\frac{Z^2}{\beta^2(T)}\frac{1}{\nu}\chi(Z,T,h\nu/T).
\end{equation}
Tabulated data from~\cite{Seltzer1985} are used for $\chi(Z,T,h\nu/T)$ in the following.

We note that the range of validity of expressions~\ref{eqn:xs-freefree} and~\ref{eqn:xs-scaled} is not firmly established, and that another expression should hold in the energy range where both expressions do not apply, namely between a few tens and a few hundreds of electron volts. In the absence of such an expression, we restrict ourselves to extrapolate both expressions outside their range of validity and select the minimum value at a fixed energy. We stress, however, that the influence of this energy range on our final results turns out to be marginal. 

These cross sections are calculated for a single target in vacuum. For low photon energies, they behave as $1/\nu$. In materials, however, these infra-red divergences are known to be cancelled by effects due to successive interactions of the electrons, namely the LPM effect, or due to the interaction of the radiated field with the electrons, namely the dielectric effects. These effects are presented in detail below.

\subsection{LPM Suppression}
\label{lpm}

The LPM effect is the suppression of photon production due to the multiple scattering of the electron~\cite{Landau1953,Migdal1956}: if an electron undergoes multiple scattering while traversing the `formation zone`, the Bremsstrahlung amplitudes from before and after the scattering can interfere, reducing the probability of Bremsstrahlung photon emission for photon energies below a certain value.  On the condition that $h\nu/(T+mc^2)<(T+mc^2)/E_{\mathrm{LPM}}$, this leads to the following suppression factor:
\begin{equation}
\label{eqn:suppression_lpm}
S_{\mathrm{LPM}}(T,\nu)=\frac{\sqrt{h\nu E_{\mathrm{LPM}}}}{T+mc^2},
\end{equation}
with $E_{\mathrm{LPM}}=\alpha m^2X_0/4hc\simeq 1.4\times 10^{17}~$eV in air. Here, $m$ is the electron mass, $c$ the speed of light, and $X_0$ the radiation length in air. In these conditions, this suppression factor becomes important at gigahertz frequencies for electron kinetic energies $T$ greater than $\simeq 1~$MeV.

\subsection{Dielectric Suppression in Air}
\label{dielectric1}

Beside the LPM effect, the Bremsstrahlung process can be suppressed by the \textit{dielectric effect}. The dielectric suppression occurs because photons coherently forward Compton scatter off the electrons of the medium in the formation zone. This introduces a phase shift into the wave function which leads, if the phase shift accumulated over the formation length is large enough, to a loss of coherence. 

The suppression factor is given by the ratio of the in-material to vacuum formation lengths of the radiated photons~\cite{Ter1972}.  The formation length $l_f$ is the size of the virtual photon exchanged between the electron and the nucleus, which is given by the uncertainty principle applied to the momentum transfer $q_\parallel$ between the electrons and the nucleus: $l_f=\hbar/q_\parallel$. For $h\nu=\hbar\omega\ll T$, this momentum transfer is given by~\cite{Ter1972}:
\begin{equation}
\label{eqn:qpar}
q_\parallel=p-p^{\prime}-\sqrt{\epsilon(\omega)}\hbar\omega/c,
\end{equation}
with $p$ and $p^\prime$ the initial and final momentum of the electron and $\epsilon(\omega)$ the \textit{dielectric coefficient} of the medium. 

To estimate the dielectric coefficient in air, we consider a widely-used simple model to estimate the current generated by the electrons bounded to the molecules of the atmosphere in presence of the radiated field. Given that at gigahertz frequencies, the wavelength of the radiated field is much larger than the distance over which the electrons oscillate, the radiated field can be considered as homogeneous in space so that in terms of the electron momentum $\mathbf{p}$, the equation of motion for $\mathbf{p}$ reads as
\begin{equation}
\frac{\dif ^2\mathbf{p}}{\dif t^2}=-\omega_0^2\mathbf{p}-ie\omega\mathbf{E},
\end{equation}
with $\mathbf{E}$ the electric field of the radiation and $\omega_0$ the binding angular frequency of the electrons in atoms, of the order of a few hundreds of terahertz. Note that the term associated to the magnetic field is neglected, since its amplitude is lower by a factor $\beta(T)$ for a radiation, with $T$ of the order of a few electron volts here. The steady-state solution in Fourier space reads as $\tilde{\mathbf{p}}=-ie\omega\tilde{\mathbf{E}}/(\omega_0^2-\omega^2)$, allowing us to estimate the current $\tilde{\mathbf{j}}$ as
\begin{equation}
\label{eqn:j}
\tilde{\mathbf{j}}(\omega)=-i\frac{ne^2\omega}{m(\omega_0^2-\omega^2)}\tilde{\mathbf{E}}(\omega),
\end{equation}
with $n$ the electron density in the atmosphere. Given that the current is related to the conductivity of the medium through $\tilde{\mathbf{j}}=\sigma\tilde{\mathbf{E}}$ and that the latter is related to the dielectric coefficient through $\epsilon(\omega)=1+i\sigma(\omega)/\epsilon_0\omega$, the dielectric coefficient in air is obtained as
\begin{equation}
\label{eqn:epsilon1}
\epsilon(\omega)=1+\frac{\omega_p^2}{\omega_0^2-\omega^2},
\end{equation}
with $\omega_p=\sqrt{ne^2/m\epsilon_0}$ the plasma pulsation. 

Inserting this expression into equation~(\ref{eqn:qpar}), the suppression factor $S_{\dif }$ is obtained as a function of the kinetic energy $T$ of the incoming electron. For $\nu$ of the order of the gigahertz, it becomes important for $T$ greater than $\simeq 0.3~$MeV, where it can be approximated as\footnote{It is to be noted that compared to the expression in~\cite{Anthony1996} for $S_{\dif }$, $\nu_0^2$ replaces $\nu^2$ in our expression. This is because, in contrast to~\cite{Anthony1996}, frequencies $\nu$ considered in this study are much smaller than the atomic binding frequency $\nu_0$.}
\begin{equation}
\label{eqn:S_d}
S_{\dif }(T)\simeq\frac{\nu_0^2}{\nu_0^2+\nu_p^2T^2}.
\end{equation}

\subsection{Total Suppression}
\label{total}

Both suppressions reduce the effective formation length of the photon, so the suppressions do not simply multiply~\cite{Galitsky1964}. For relativistic electrons, the total suppression factor $S$ can be shown to be~\cite{Galitsky1964}
\begin{equation}
\label{eqn:suppression}
S(T,\nu)=\frac{1}{2}\left(\sqrt{(1+1/S_{\dif })^2S_{\mathrm{LPM}}^4+4S_{\mathrm{LPM}}^2}-\left(1+1/S_{\dif }\right)S_{\mathrm{LPM}}^2\right).
\end{equation}

\begin{figure}[!t]
\centering
\includegraphics[width=12cm]{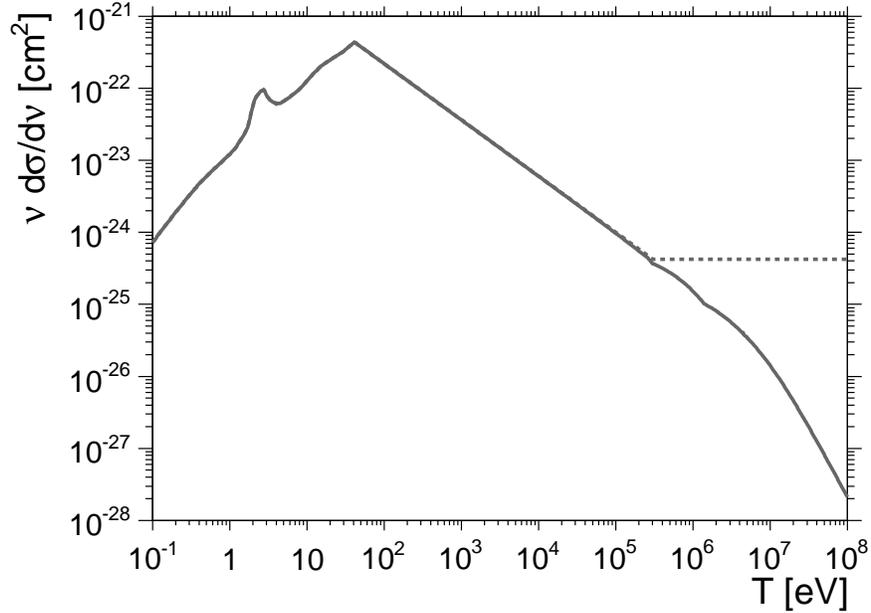}
\caption{\small{Bremsstrahlung differential cross section multiplied by $\nu$ as a function of the kinetic energy of the electron, for a nitrogen molecule target.}}
\label{fig:xs}
\end{figure}

The resulting Bremsstrahlung differential cross section $\dif\sigma/\dif\nu$, accounting for suppression effects in air, is shown in figure~\ref{fig:xs}, multiplied by the frequency $\nu$, as a function of the kinetic energy of the incoming electron. This parameterization for  $\nu\dif\sigma/\dif\nu$ pertains to a frequency range for the photons such that $h\nu\ll T$. Without the suppression effects, $\nu\dif\sigma/\dif\nu$ would be almost constant above a few hundreds of kilo-electron volts, as shown by the dotted line.

\subsection{Plasma Absorption Effects}
\label{dielectric2}

In addition to dielectric effects in air, dielectric effects in the plasma of free electrons and neutral molecules can also enter into play, depending on the density of free electrons. The charge density induced by these electrons is the source of an electric field which influences the motion of the electrons and the conductivity of the plasma to radiation fields. In general, depending on the geometry of the considered plasma, both the conductivity $\underline{\sigma}$ and the dielectric coefficient $\underline{\epsilon}$ are tensors of complex numbers. This implies an absorption of the radiated field such that each component of the absorption length vector $\mathbf{L}(\omega)$ is
\begin{equation}
\label{eqn:absorption}
L_i(\omega)=\frac{c}{2\omega}\frac{1}{\mathrm{Im}\left(\sqrt{[\epsilon(\omega)]_{ii}}\right)}.
\end{equation}
The estimation of the dielectric coefficient related to the plasma is thus of central importance in estimating whether the radiated field can be observed or not at some distance of the emission point. The plasmas that will be considered in next sections consist of the neutral molecules in air and of the numerous low-energy ionization electrons left in air after the passage of high-energy electrons. The dielectric tensors will be estimated in sections~\ref{dielectric-shower} and~\ref{comments} for each considered geometry of the plasmas, to determine the absorption lengths.

\section{Molecular Bremsstrahlung Radiation in Extensive Air Showers}
\label{mbr1}

In this section, we consider a description of the high-energy electrons\footnote{Hereafter, the term \textit{electrons} stands for both electrons and positrons.} in the electromagnetic cascade, providing only gross properties. This description will allow us, however, to derive a realistic order of magnitude of the spectral intensities that can be expected at ground level from molecular Brems\-strah\-lung radiation by these electrons and by the numerous low-energy ionization electrons left in the atmosphere along the shower track. To facilitate comparisons with previous works, the parameters of this description will be tuned to apply to a vertical shower with primary energy $E=10^{17.5}~$eV. 

\subsection{High-Energy Electrons in Extensive Air Showers}
\label{highener}

An extensive air shower is hereafter considered as a thin plane front of high-energy charged particles propagating in the atmosphere at the speed $c$. After the succession of a few initial steps in the cascade, all showers can be roughly described in terms of reproducible macroscopic states, and the longitudinal development of the electromagnetic cascade depends only on the cumulated slant depth $X$ in a universal way except for a translation depending logarithmically on $E$ and for a global factor roughly linear in $E$. In particular, for any given slant depth $X$, or equivalently any altitude $a$, the total number of primary electrons, $N_{\mathrm{e}}$, can be adequately parameterized by the Gaisser-Hillas function as~\cite{GH}:
\begin{equation}
\label{eqn:gh}
N_{\mathrm{e}}(a)=N_{\mathrm{max}}\left(\frac{X(a)-X_0}{X_{\mathrm{max}}-X_0}\right)^{\frac{X_{\mathrm{max}}-X_0}{\lambda}}\exp{\left(\frac{X_{\mathrm{max}}-X(a)}{\lambda}\right)},
\end{equation}
with $X(a)$ the depth corresponding to the altitude $a$, $X_0$ the depth of the first interaction, $X_{\mathrm{max}}$ the depth of shower maximum, $N_{\mathrm{max}}$ the number of particles observed at $X_{\mathrm{max}}$, and $\lambda$ a parameter describing the attenuation of the shower.

Of relevant importance for the following is the way the electrons are distributed within the shower in terms of energy $T$, radial distance to the shower axis $r$ and angles around the shower axis $\Omega=(\chi,\psi)$. Based on Monte-Carlo simulations of showers initiated by hadrons, parameterizations have been obtained in~\cite{Lafebre2009} by factoring the energy spectrum, the energy-dependent angular spectrum, and the energy-dependent radial spectrum expressed in term of the \textit{relative evolution stage} $(X-X_{\mathrm{max}})/X_0$ of the showers. For a vertical shower, the dependence in relative evolution stage reflects into a dependence in altitude $a$ only. 

\begin{figure}[!t]
\centering
\includegraphics[width=12cm]{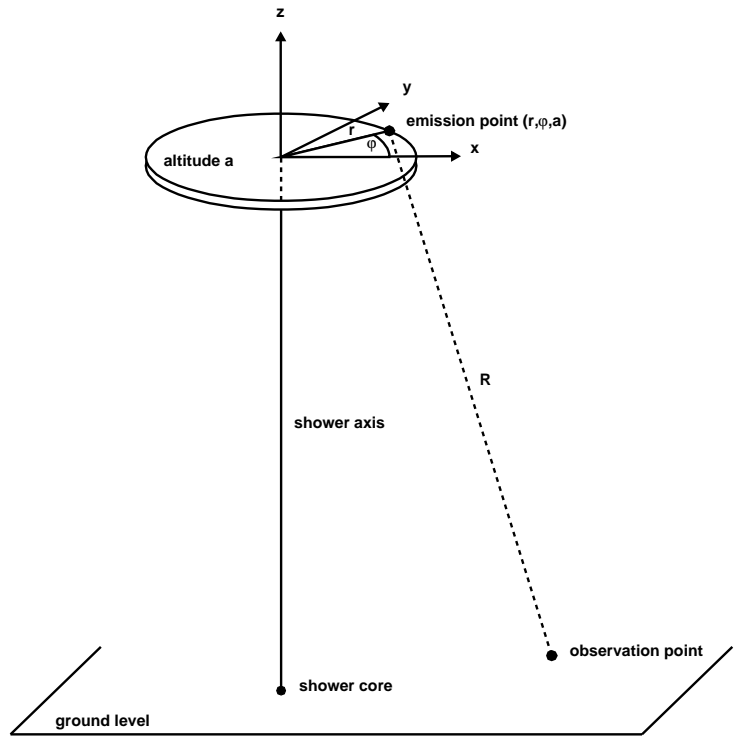}
\caption{\small{Geometry of a vertical shower used throughout the paper.}}
\label{fig:shower}
\end{figure}

For convenience, we express the longitudinal dependences in terms of $a$ hereafter. The relative number of electrons per energy band along the shower can be accurately described by the following distribution:
\begin{equation}
\label{eqn:f_Te}
f_{T}(T,a)\propto \frac{T^{\gamma_1}}{(T+\epsilon_1)^{\gamma_1}(T+\epsilon_2)^{\gamma_2}}.
\end{equation}
The parameters $\epsilon_1$, $\epsilon_2$, $\gamma_1$ and $\gamma_2$ depend on $X(a)$ and are taken from~\cite{Lafebre2009}.  Around $X_{\mathrm{max}}$, this distribution has a maximum for $T\simeq20~$MeV. Besides, the lateral extension of the electromagnetic cascade depends on the mean free path of the particles and can be expressed in terms of the \textit{Moli\`ere radius} $R_{\mathrm{M}}$ such that 90\% of the energy is contained within a distance $r$ from the axis such as $r<R_{\mathrm{M}}$.  This lateral extension can be described by the Nishimura-Kamata-Greisen (NKG) function known to reproduce reasonably well the observations~\cite{}. To reproduce the main features of the energy dependences of this lateral extension, the shape of the NKG function is used but with different parameters:
\begin{equation}
\label{eqn:f_R}
f_{R}(r,T,a)\propto \left(\frac{r}{R_{\mathrm{M}}}\right)^{\zeta_0^\prime}\left(\mu_1^\prime+\frac{r}{R_{\mathrm{M}}}\right)^{\zeta_1^\prime}.
\end{equation}
Differently from the traditional NKG function, the parameter $\mu_1^\prime$ depends here on the energy. $\zeta_1^\prime$ depends on the relative evolution stage, while $\zeta_0^\prime$ depends also on the energy in addition to the standard dependence on the relative evolution stage in the  function. The exact parameterizations are taken from~\cite{Lafebre2009}. Finally, the angular spectrum of the electrons reflects the collimation around the shower axis, the collimation being more important for more energetic electrons. It can be accurately parameterized as:
\begin{equation}
\label{eqn:f_Omega}
f_{\Omega}(\chi,\psi,T,a)\propto \left[\left(\mathrm{e}^{b_1}\chi^{\alpha_1}\right)^{-1/\sigma}+\left(\mathrm{e}^{b_2}\chi^{\alpha_2}\right)^{-1/\sigma}\right]^{-\sigma}.
\end{equation}
The energy dependences of $b_1$, $\alpha_1$, $b_2$ and $\alpha_2$ are taken from~\cite{Lafebre2009}. The parameter $\sigma$ describes the smoothness of the transition from the first term of importance near the shower axis to the second term being relevant further away. 

In our work, all these distributions are normalized simultaneously:
\begin{equation}
\label{eqn:normalization}
2\pi\int\dif T\int\dif \Omega\int\dif r~rf_{T}(T,a)f_{R}(r,T,a)f_{\Omega}(\chi,\psi,T,a)=1.
\end{equation}
In this way, the number of primary $e^+/e^-$ per unit surface, per energy band and per solid angle, $n_{\mathrm{e}}(r,a,T,\chi,\psi)$, is simply obtained by folding the $f$ distributions to the Gaisser-Hillas function:
\begin{equation}
\label{eqn:ne}
n_{\mathrm{e}}(r,a,T,\chi,\psi)=N_{\mathrm{e}}(a)~f_{T}(T,a)f_{R}(r,T,a)f_{\Omega}(\chi,\psi,T,a).
\end{equation}
Also, the number of primary $e^+/e^-$ per unit surface, $n_{\mathrm{e}}^{\mathrm{NKG}}(r,a,T,\chi,\psi)$, is obtained by marginalizing over $T$ and $(\chi,\psi)$:
\begin{equation}
\label{eqn:ne_}
n_{\mathrm{e}}^{\mathrm{NKG}}(r,a)=N_{\mathrm{e}}(a)~\int\dif T\int\sin\chi\dif\chi\int\dif\psi f_{T}(T,a)f_{R}(r,T,a)f_{\Omega}(\chi,\psi,T,a).
\end{equation}
This corresponds then to the NKG parameterization.

The instantaneous \textit{flux} $\phi_{\mathrm{e}}(r,a,T,\chi,\psi,t)$ of electrons per energy band and per solid angle is the main quantity of interest in estimating the production rate of photons per volume unit through molecular Bremsstrahlung. Under the approximation that all particles are contained within the infinitely thin shower front moving at the speed $\simeq c$, the instantaneous flux of electrons crossing any elementary surface corresponds to the number of electrons per surface unit $n_{\mathrm{e}}$ when these electrons are located in $(r,a)$ at time $t$. The differential flux of electrons 
$\phi^\Omega_{\mathrm{e}}(r,a,T,\chi,\psi,t)$ per energy band is then:
\begin{equation}
\label{eqn:fluxelec0}
\phi^\Omega_{\mathrm{e}}(r,a,T,\chi,\psi,t)\simeq n_{\mathrm{e}}(r,a,T,\chi,\psi)~\delta(t-t_{\mathrm{sf}}(r,a)),
\end{equation}
where the Dirac function\footnote{Note that the structure of the shower front is actually much more complex than the ideal picture given here in terms of an infinitely thin front. However, we have checked that the inclusion of the time and space structure of the front, as presented in~\cite{Lafebre2009}, does not lead to significant changes for the results presented in this study. Hence, for convenience, we defer to the Dirac function which leads to several simplifications in the formulas.} is introduced to guarantee the presence of the particles in $(r,a)$ at time $t$ with the passage of the shower front at time $t_{\mathrm{sf}}(r,a)$. For a vertical shower, and considering that $t=0$ when the shower hits the ground, $t_{\mathrm{sf}}$ depends only on $a$ and is simply $t_{\mathrm{sf}}(a)=-a/c$.

\subsection{Low-Energy Ionization Electrons}
\label{lowener}

Through the passage of charged particles in the atmosphere, the energy of an extensive air shower is deposited mainly through the ionization process. For one single high-energy electron travelling over an infinitesimal distance $\dif x$, and for a mass density $\rho_m$ of molecular nitrogen or oxygen, the average number of ionization electrons per unit length and per kinetic energy band can be estimated as:
\begin{equation}
\label{eqn:d2N_dadT}
\frac{\dif ^2N_{\mathrm{e,i}}}{\dif x~\dif T}(T)=\frac{\rho_m~f_0(T)}{I_0+\left\langle T\right\rangle}\left\langle\frac{\dif E}{\dif X}\right\rangle,
\end{equation}
with $I_0$ the ionization potential to create an electron-ion pair in air, and the bracketed expression $\left\langle\dif E/\dif X\right\rangle$ stands for the mean energy loss of primary electrons per grammage unit.  The distribution in kinetic energy of the resulting ionization electrons is described here by the normalized function $f_0(T)\equiv f(T,t=0)$. This distribution has been experimentally determined and accurately parameterized for primary electrons with kinetic energies $T^{\mathrm{p}}$ up to several kilo-electron volts~\cite{Opal}. For higher kinetic energies $T^{\mathrm{p}}$, relativistic effects as well as indistinguishability between primary and secondary electrons have been shown to modify the low-energy behaviour~\cite{Arqueros}. To account for these effects, we adopt the analytical expression provided in~\cite{Arqueros2}:
\begin{equation}
\label{eqn:f0}
f_0(T)=\frac{8\pi ZR_y^2}{m\left(\beta(T^{\mathrm{p}})c\right)^2}\frac{1+C\exp{(-T/T_\mathrm{k})}}{T^2+\overline{T}^2},
\end{equation}
where $T$ ranges from 0 to $T^{\mathrm{max}}=(T^p-I_0)/2$ due to the indistinguishability between primary and secondary electrons, the constant $C$ is determined in the same way as in~\cite{Opal} so that $\int \mathrm{d}T~f_0(T)$ reproduces the total ionization cross section, $T_\mathrm{k}=77~$eV is a parameter acting as the boundary between close and distant collisions, and $\overline{T}$ a measured parameter such that $\overline{T}=13.0~(17.4)~$eV for nitrogen (oxygen). In the energy range of interest, this expression leads to $\left\langle T\right\rangle\simeq 40~$eV, in agreement with the well-known stopping power. The instantaneous number of ionization electrons per unit volume and per kinetic energy band is then obtained by coupling equation~(\ref{eqn:d2N_dadT}) to the number of high-energy charged particles per surface unit:
\begin{equation}
\label{eqn:nei}
n_{\mathrm{e,i}}(r,a,T,t=0)=\frac{\rho_m(a)f_0(T)}{I_0+\left\langle T\right\rangle}~\left\langle\frac{\dif E}{\dif X}\right\rangle~n_{\mathrm{e}}^{\mathrm{NKG}}(r,a).
\end{equation}
Here, the energy dependence of the number of high-energy charged particles in the shower, as expressed in equation~(\ref{eqn:ne}), can be ignored due to the constancy of $\left\langle\dif E/\dif X\right\rangle$ in the energy range of interest defined by $f_{T}$. As well, the low energy of the produced ionization electrons makes the angular dependence of their initial momentum irrelevant to consider for the following. This is because they can be considered as quasi-static in space to a good approximation and consequently as point sources of gigahertz photons, photons which are expected to be emitted \textit{isotropically} from equation~(\ref{eqn:dsigma_ff}). Hence, only $n_{\mathrm{e}}^{\mathrm{NKG}}(r,a)$ is relevant to consider to obtain $n_{\mathrm{e,i}}(r,a,T,t=0)$. From equation~(\ref{eqn:nei}), the instantaneous differential \textit{flux} $\phi^\Omega_{\mathrm{e,i}}(r,a,T,\chi,\psi)$ of ionization electrons per kinetic energy band is obtained in the same way as 
\begin{equation}
\label{eqn:flux_nei}
\phi^\Omega_{\mathrm{e,i}}(r,a,T,t=0,\chi,\psi)=\frac{c\beta(T)f_0(T)}{4\pi(I_0+\left\langle T\right\rangle)}\left\langle\frac{\dif E}{\dif X}\right\rangle\rho_m(a)~n_{\mathrm{e}}^{\mathrm{NKG}}(r,a).
\end{equation}

The distribution function $f_0(T)$ describes the kinetic energy dependence of the ionization electrons at their time of creation. The time evolution of $\phi^\Omega_{\mathrm{e,i}}(r,a,T,\chi,\psi)$ is then fully encompassed in the time dependence of $f(T,t)$. This distribution can be related to the distribution in velocities $f_V(v,t)$ through a jacobian transformation, the latter distribution being obtained by marginalising the three dimensional distribution $f_\mathbf{V}(\mathbf{v},t)$. $f_\mathbf{V}(\mathbf{v},t)$ is governed by the following Boltzmann equation accounting for the interactions of the electrons with electromagnetic fields and the collisions of the electrons on target molecules considered at rest:
\begin{equation}
\label{eqn:boltzmann_general}
\frac{\partial f_\mathbf{V}(\mathbf{v},t)}{\partial t}+\frac{e}{m}\left(\mathbf{E}+\mathbf{v}\times\mathbf{B}\right)\cdot\mathbf{\nabla}_{\mathbf{v}}f_\mathbf{V}(\mathbf{v},t)=-\nu_c(v)f_\mathbf{V}(\mathbf{v},t)+\sum_m\frac{n_m}{4\pi v^2}\int\dif \mathbf{v}'~v'\frac{\dif \sigma^m(v',v)}{\dif v}f_\mathbf{V}(\mathbf{v}',t),
\end{equation}
where $\nu_c(v)$ denotes the collision rate accounting for the sources of depletion of electrons at velocity $v$, and where the last term includes all contributions to the regeneration of the distribution function at the same velocity. Although the distribution $f_\mathbf{V}$ can depend on the position in space through the corresponding dependence of the electromagnetic field and/or the density of target molecules, the spatial dispersion term in the Boltzmann equation is neglected. This is because, given their low energy and their rate of disappearance that we anticipate (see below) at the level of at most a few hundreds of nanoseconds, ionization electrons remain confined within a few $10^{-9}$~m$^3$ and can thus be considered to a good approximation as quasi-static in space. 

To solve equation~(\ref{eqn:boltzmann_general}) in the context of extensive air showers, we consider that the contribution of the electromagnetic term is small compared to the one of the collision term and thus neglect initially. Later, the electromagnetic term will be re-inserted in the solution as a small perturbation. The impact of this perturbation will be quantified to check the range of validity of the solution in terms of the amplitude of the electromagnetic field. 

\begin{figure}[!h]
\centering
\includegraphics[width=12cm]{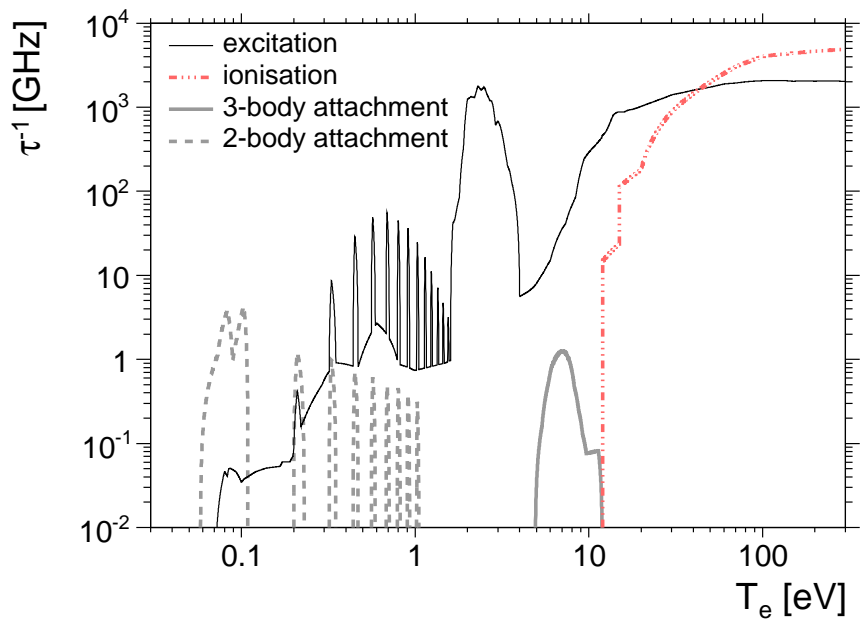}
\caption{\small{Dominant collision rates of low-energy electrons in air, at sea level for a water concentration of 3,000 ppm.}}
\label{fig:collision_rate}
\end{figure}

Hence, neglecting the electromagnetic term and for isotropic velocities at $t=0$, the function $f_{\mathbf{V}}$ does not depend on the direction of the velocity vector $\mathbf{v}$ so that the equation reduces to the one-dimensional distribution $f_v(v,t)=4\pi v^2f_{\mathbf{V}}(\mathbf{v},t)$. Since the cross sections of the processes of interest are generally given in terms of kinetic energy, it is more convenient to write the corresponding Boltzmann equation in terms of $T$:
\begin{equation}
\label{eqn:boltzmann_1d}
\frac{\partial f(T,t)}{\partial t}=-\nu_c(T)f(T,t)+\sum_m n_mc\int\dif T^\prime~\frac{\dif \sigma^m(T^\prime,T)}{\dif T}\beta(T^\prime)f(T^\prime,t).
\end{equation}
The relevant collision rates $\nu_i=n_mc\beta(T)\sigma^m_i(T)$, all taken from experimental tabulated data in~\cite{JILA}, are shown in figure~\ref{fig:collision_rate} as a function of the kinetic energy. Going down in energy, the main features of the different rates can be highlighted in the following way:
\begin{itemize}
\item For $T\geq 40~$eV, ionization on N$_2$ and O$_2$ molecules is the dominant process, causing energy losses on a time scale below the picosecond.
\item For 4~eV~$\leq T\leq 40~$eV, excitation on electronic levels of N$_2$ and O$_2$ molecules is the dominant process.  The corresponding energy losses occur on time scales going from picoseconds to a few nanoseconds when going down in energy.
\item For 1.7~eV~$\leq T\leq 4~$eV, resonances for excitation on N$_2$ and O$_2$ molecules through ro-vibrational processes cause energy losses on a time scale of the picosecond.
\item For 0.2~eV~$\leq T\leq 1.7~$eV: 
\begin{itemize}
\item Resonances for excitation on N$_2$ and O$_2$ molecules through ro-vibrational processes quantized in energies are visible through the peaks in continuous line. The corresponding energy losses occur on a time scale of a few tens of picoseconds.
\item Resonances for two-body attachment process on O$_2$ molecules quantized in energies are visible through the peaks in dashed line. The corresponding time scale of disappearance of the electrons is of the order of the nanosecond. This process is subdominant compared to the previous one. 
\item For energies between the quantized ones where ro-vibrational and attachment processes occur, energy losses on the small fractions of CO$_2$ and H$_2$O molecules are important to consider. The corresponding energy losses occur on a time scale of a nanosecond.
\end{itemize}
\item For $T\leq 0.2~$eV, energy losses on CO$_2$ and H$_2$O molecules degrade electron energies down to 0.1~eV, where the two-body attachment process make them disappearing on a time scale of a few nanoseconds.
\item Although the abundance of Ar in air is larger than for CO$_2$ and H$_2$O mole\-cu\-les, the corresponding energy losses occur at energies above 1~eV, and are thus negligible compared to the energy losses due to collision on N$_2$ and O$_2$ molecules.
\item Other processes such as Bremsstrahlung or recombination on O$_2^+$ ions have rates much smaller, and they are negligible in determining the evolution with time of the kinetic energy distribution of the electrons.
\end{itemize}
The concentration of H$_2$O molecules in the atmosphere is subject to large variations. In this study, we use typical values ranging from 3,000~ppm to 10,000~ppm to probe the systematic impact of this concentration on the final result. Besides, note that in reference~\cite{AlSamarai2015}, the parameterization used for the two-body attachment process was taken from an older work compared to data available in~\cite{JILA}. In particular, the quantization of the rates in energy was absent. In addition, energy losses for $T$ below 1~eV due to collision on CO$_2$ and H$_2$O molecules were not accounted for due to the smaller fractions of these molecules with respect to N$_2$ and O$_2$. The quantization of the  two-body attachment process forces us, however, to consider interactions with these molecules. Otherwise, electrons with energies between two quantized peaks of attachment would be trapped much longer and the photon emission from these electrons would be highly overestimated. 

\begin{figure}[!t]
\centering
\includegraphics[width=12cm]{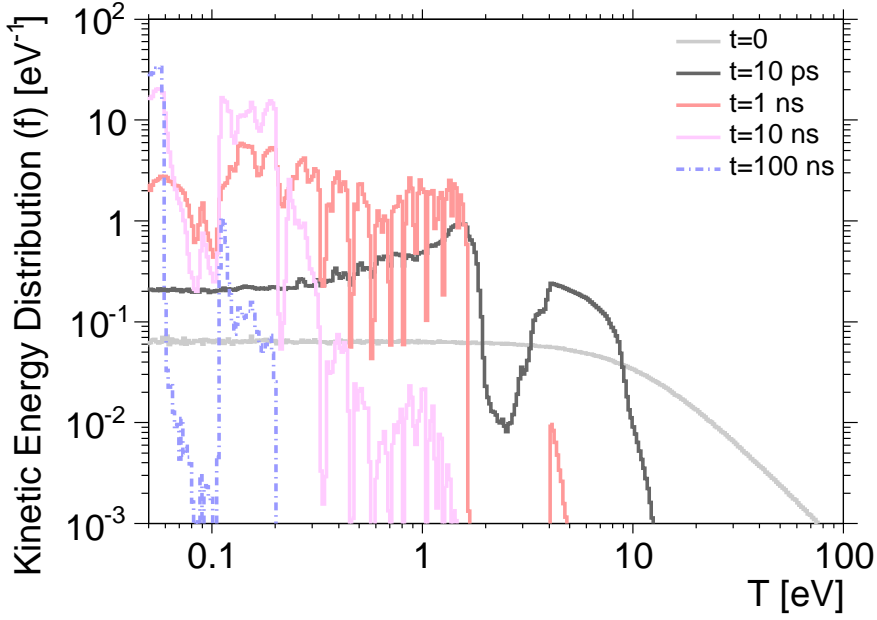}
\caption{\small{Distribution function ($f$) in kinetic energy of the ionization electrons left along the shower track in the atmosphere after different times, at sea level for a water concentration of 3,000 ppm.}}
\label{fig:f}
\end{figure}

With these ingredients, the Boltzmann equation governing the time evolution of the $f$ function reads as:
\begin{eqnarray}
\label{eqn:boltzmann}
\frac{\partial f}{\partial t}(T,t,a)&=&-\sum_{m=\mathrm{N}_2,\mathrm{O}_2}\hspace{-3mm}n_{m}(a)c\beta(T)\left(\sigma^{m}_{\mathrm{ion}}(T)+\sigma^{m}_{\mathrm{exc}}(T)\right)f(T,t) \nonumber \\
&-&\hspace{-5mm}\sum_{m=\mathrm{CO}_2,\mathrm{H}_2\mathrm{O}}\hspace{-5mm}n_{m}(a)c\beta(T)\sigma^{m}_{\mathrm{exc}}(T)f(T,t)  - n_{\mathrm{O}_2}(a)c\beta(T)\sigma^{\mathrm{O}_2}_{\mathrm{att}}(T)f(T,t) \nonumber \\
&+&\hspace{-3mm}\sum_{m=\mathrm{N}_2,\mathrm{O}_2} \hspace{-3mm}n_m(a)c\int_{T}^{T^{\mathrm{max}}}\dif T^\prime\beta(T^\prime)\left(\frac{\dif \sigma^m_{\mathrm{ion}}}{\dif T}(T^\prime, T)+\frac{\dif \sigma^m_{\mathrm{ion}}}{\dif T}(T^\prime, T^\prime-T)\right)f(T^\prime,t) \nonumber \\
&+&\hspace{-3mm}\sum_{m=\mathrm{N}_2,\mathrm{O}_2,\mathrm{CO}_2,\mathrm{H}_2\mathrm{O}}\hspace{-5mm} n_m(a)c\int_{T}^{T^{\mathrm{max}}}\dif T^\prime\beta(T^\prime)\frac{\dif \sigma^m_{\mathrm{exc}}}{\dif T}(T^\prime, T)f(T^\prime,t),
\end{eqnarray}
where $\sigma^m_i$ denotes the cross sections of interest, namely ionization, excitation of electronic levels (including ro-vibrational excitation) and attachment processes for a mole\-cu\-le $m$. The first three terms in the right hand side stand for the disappearance of electrons with kinetic energy $T$, while the last two terms stand for the appearance of electrons with kinetic energy $T$ due to ionization and excitation reactions initiated by electrons with higher kinetic energy $T^\prime$. Note that in the case of ionization, a second electron emerges from the collision with kinetic energy $T^\prime-T$. 

The time evolution of the $f$ function is illustrated in figure~\ref{fig:f} at different times and at sea level, as obtained by solving equation~(\ref{eqn:boltzmann}) by Monte-Carlo. Dips at kinetic energies corresponding to collision rates at quantized energies depicted in figure~\ref{fig:collision_rate} are clearly seen. After a time of 1~ns, almost all electron energies are already below the ionization threshold. After 10~ns, the number of electrons with energies around 1~eV drops off to a large extent, due to excitation reactions. After 100~ns, only a population of very low energy electrons remains, a population that will be irrelevant for the following. With high accuracy, it turns out that the $f$ distribution can be parameterized as $f(T,t)=f_0(T)\exp{(-\xi(T,t))}$, with $\xi(T,t=0)=0$. In other words, and this will be convenient in the following, the right hand side of equation~(\ref{eqn:boltzmann}) can be replaced by a single effective term $-\nu_{\mathrm{eff}}(T,t)f(T,t)$, with $\nu_{\mathrm{eff}}=\partial \xi/\partial t$. $\nu_{\mathrm{eff}}$ is to be interpreted as an \textit{effective collision rate}, which can be negative in some kinetic energy and time ranges.

Having determined a convenient parameterization of equation~(\ref{eqn:boltzmann}) and of its solution, we proceed now with the electromagnetic term neglected so far in equation~(\ref{eqn:boltzmann_general}). Neglecting the electromagnetic field induced by radiations, we consider the field induced by the numerous low-energy ionization electrons only. Since the ionization electrons are quasi-static in space, the magnetic term can be neglected compared to the electric one. Considering this latter term as a perturbation of the previous solution, denoted hereafter by $f_\mathbf{V}^{(0)}$, once expressed in terms of velocities, we decompose the solution of the global equation such that $f_\mathbf{V}=f_\mathbf{V}^{(0)}+f_\mathbf{V}^{(1)}$. Neglecting the product of small quantities gives to first order the following equation for $f_\mathbf{V}^{(1)}$:
\begin{equation}
\label{eqn:boltzmann_perturb}
\frac{\partial f^{(1)}_\mathbf{V}(\mathbf{v},t)}{\partial t}+\frac{e}{m}\mathbf{E}\cdot\mathbf{\nabla}_{\mathbf{v}}f^{(0)}_\mathbf{V}(\mathbf{v},t)=-\nu_\mathrm{eff}(v,t) f^{(1)}_\mathbf{V}(\mathbf{v},t).
\end{equation}
Given the geometry of the simplified vertical shower considered in this study, and depicted in figure~\ref{fig:shower}, the electric field induced by the ionization electrons is independent of the azimuthal angle. In this case, the solution for $f^{(1)}_\mathbf{V}$ is:
\begin{equation}
\label{eqn:sol_perturb}
f^{(1)}_\mathbf{V}(\mathbf{v},t)=-\frac{e}{m}\int_0^t\dif t'~\left(E_r(t')\frac{\partial f_\mathbf{V}^{(0)}(\mathbf{v},t')}{\partial v_r}+E_z(t')\frac{\partial f_\mathbf{V}^{(0)}(\mathbf{v},t')}{\partial v_z}\right)\exp{\left(\xi(v,t')-\xi(v,t)\right)}.
\end{equation}
The impact of this perturbation on the photon emission by the electrons will be quantified in next subsection.

\subsection{Spectral Intensity at Ground from Bremsstrahlung}
\label{sintens}

From the photon production rate $r_\gamma$, the emitted power per unit volume and under each direction $(\chi,\psi)$ can be obtained by coupling the production rate to the energy of the emitted photons, so that the emitted spectral power per volume unit can be written as:
\begin{equation}
\label{eqn:spectral_power_1}
\frac{\dif ^2P}{\dif \nu \dif V}(r,a,\chi,\psi,\nu)=h\nu r_\gamma(r,a,\chi,\psi,\nu).
\end{equation}
At any distance $R$ from the emission point, the spectral intensity traversing any surface element $\dif S=R^2\dif \Omega$ is equal to the spectral power emitted within $\dif \Omega$ attenuated by an exponential factor $\exp{(-\kappa(R,\nu))}$ to account for an eventual absorption induced by the dielectric effects of the plasma. Here, $\kappa(R)$ is an `optical` depth such that $\kappa(R,\nu)=\int_0^R \dif \mathbf{s}\cdot\mathbf{L}^{-1}(\nu,\mathbf{s})$. In this way, the observable spectral intensity at any ground position, $\Phi_{\mathrm{g}}$, is simply the sum of the uncorrelated contributions of the individual encounters weighted by $R^{-2}$, with $R$ the distance between the position at ground $\mathbf{x}_{\mathrm{g}}$ and the position of the current source in the integration:
\begin{equation}
\label{eqn:spectral_intensity_0}
\Phi_{\mathrm{g}}(\mathbf{x}_{\mathrm{g}},t,\nu)=\int_0^\infty r\dif r\int_0^{2\pi}\dif \varphi\int_0^\infty \dif a~\frac{\exp{(-\kappa(R,\nu))}}{R^2(r,\varphi,a)}\frac{d^2P}{\dif \nu\dif V}(r,a,\chi(r,a),\psi(r,a),t_{\dif },\nu).
\end{equation}
Here, $t_{\dif }$ is the \textit{delayed} time at which the emission occurred:
\begin{equation}
\label{eqn:delay}
t_{\dif }\equiv t_{\dif }(t,r,\varphi,a,\nu)=t-\frac{R(r,\varphi,a)n(a,\nu)}{c},
\end{equation}
with $n(a,\nu)$ the refractive index integrated along the line of sight between the emission point and the observer. 

\begin{figure}[!t]
\centering
\includegraphics[width=12cm]{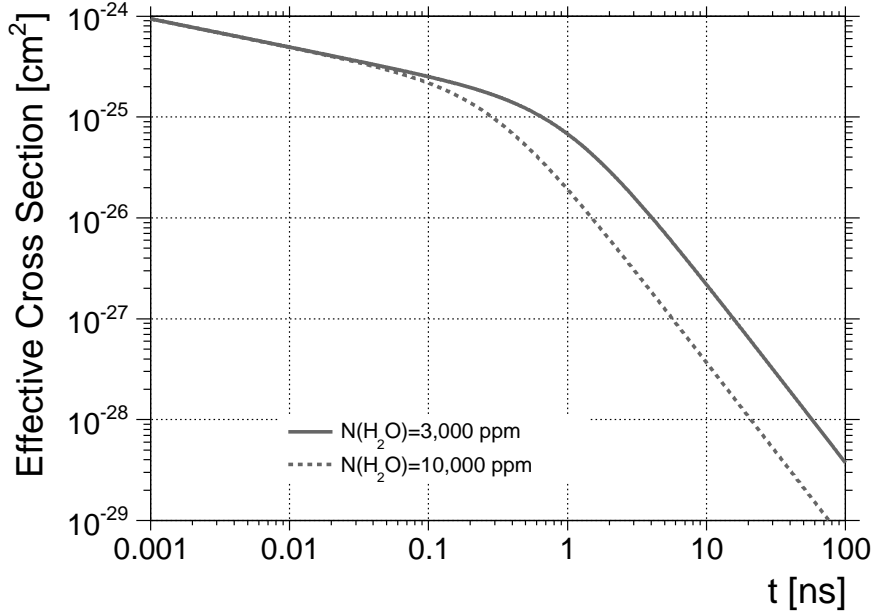}
\caption{\small{Effective cross section for the ionization electrons to produce photons by Bremsstrahlung as a function of time, at sea level (see equation~(\ref{eqn:sigmatilde})).}}
\label{fig:sigmatilde}
\end{figure}

The total spectral intensity, $\Phi_{\mathrm{g}}^{\mathrm{tot}}$, is the sum of the one from high-energy electrons of the shower, $\Phi_{\mathrm{g}}^{\mathrm{e}}$, and of the one from low-energy ionization electrons, $\Phi_{\mathrm{g}}^{\mathrm{e,i}}$. The first one is obtained by inserting equation~(\ref{eqn:fluxelec0}) into equation~(\ref{eqn:rate}), and then by inserting the resulting spectral power per volume unit into equation~(\ref{eqn:spectral_intensity_0}). Carrying out the integration over the altitude $a$ thanks to the Dirac function, it reads as:
\begin{equation}
\label{eqn:spectral_intensity_1}
\Phi_{\mathrm{g}}^{\mathrm{e}}(\mathbf{x}_{\mathrm{g}},t,\nu)=\frac{hc\mathcal{N}_A}{A}\int r\dif r\int\dif \varphi\int\dif T\int\dif \Omega^\prime~\frac{\rho_m n_{\mathrm{e}} \exp{(-\kappa(R,\nu))}}{\left|\frac{a_{\dif }n(a_{\dif })}{R}+\frac{R\partial n(a_{\dif })}{\partial a}-1\right|R^2}\nu(\nu_0,T)\frac{\dif ^2\sigma(T,\nu_0,\omega)}{\dif \nu_0\dif \Omega},
\end{equation}
with $\rho_m\equiv \rho_m(a_{\dif })$, $n_{\mathrm{e}}\equiv n_{\mathrm{e}}(r,a_{\dif },T,\chi^\prime,\psi^\prime)$, $R\equiv R(r,\varphi,a_{\dif })$ and $a_{\dif }\equiv a_{\dif }(r,\varphi,t)$ the solution of:
\begin{equation}
\label{eqn:a_d}
a_{\dif }-R(r,\varphi,a_{\dif })n(a_{\dif },\nu)=ct.
\end{equation}
Note that the frequency $\nu_0$ entering into the integrand is related to $\nu$ through $\nu=\nu_0\sqrt{1-\beta^2(T)}/(1-\beta(T))$, as a consequence of the Doppler effect due to the relativistic motion of the high-energy electrons forming the shower front. On the other hand, the second term $\Phi_{\mathrm{g}}^{\mathrm{e,i}}$ can be expressed after the solid angle integration thanks to the isotropy of the double differential cross section in the energy range of interest:
\begin{equation}
\label{eqn:spectral_intensity_2}
\Phi_{\mathrm{g}}^{\mathrm{e,i}}(\mathbf{x}_{\mathrm{g}},t)=\frac{hc\mathcal{N}_A}{8\pi A(I_0+\left\langle T\right\rangle)}\left\langle\frac{\dif E}{\dif X}\right\rangle\int r\dif r\int\dif \varphi\int \dif a~\frac{\rho_m^2n_{\mathrm{e}}^\mathrm{NKG}\exp{(-\kappa(R,\nu))}}{R^2(r,\varphi,a)}~\tilde{\sigma}({t_{\dif },a})~\Theta(t_{\dif }),
\end{equation}
with $\rho_m\equiv\rho_m(a)$, $n_{\mathrm{e}}^{\mathrm{NKG}}\equiv n_{\mathrm{e}^{\mathrm{NKG}}}(r,a)$, $\Theta$ the Heaviside function and $\tilde{\sigma}(t,a)$ is an effective cross-section shown in figure~\ref{fig:sigmatilde} defined as: 
\begin{equation}
\label{eqn:sigmatilde}
\tilde{\sigma}(t,a)=\int_0^{T^{\mathrm{max}}} \hspace{-5mm}\dif T f(T,t,a)\beta(T)\frac{\nu~\dif \sigma}{\dif \nu}(T)+\frac{2\pi}{c}\int_{0}^{v_{\mathrm{max}}}\hspace{-5mm}\dif v_r\int_{-v_z^{\mathrm{max}}}^{v_z^{\mathrm{max}}}\hspace{-5mm}\dif v_z~v_r\sqrt{v_r^2+v_z^2}f^{(1)}_{\mathrm{V}}(\mathbf{v},t)\nu\frac{\dif \sigma}{\dif \nu},
\end{equation}
with $v_z^{\mathrm{max}}=\sqrt{v_{\mathrm{max}}^2-v_r^2}$. This effective cross section governs the intensity and time dependence of the emission. It reaches $\simeq 1.7\times10^{-24}~$cm$^2$ at $t=0$ and decreases with time. The decrease is faster for a higher concentration of H$_2$O in the atmosphere, as shown in figure~\ref{fig:sigmatilde}.

Due to the general methodology adopted to get at equation~(\ref{eqn:sigmatilde}), this expression is valid as long as the second term in the right hand side does not dominate over the first one. This is the case as long as the radial component of the electric field\footnote{The term proportional to the longitudinal component of the electric field in $f_\mathbf{V}^{(1)}$ leads to 0 since the corresponding integrand is odd in $v_z$.} is less than $\simeq 100~$V~m$^{-1}$. To check the validity of the initial assumptions, we start by estimating the electric field generated by the ionization electrons from the general electric potential $V$ as obtained from a distribution of charges independent of the azimuthal angle:
\begin{equation}
\label{eqn:elec_potential}
V(r,a)=\frac{e}{4\pi\epsilon_0}\int_0^{\infty}\dif r'\int_{-ct_{\mathrm{sf}}(a)}^{\infty}\dif a'\int_0^{2\pi}\dif \varphi \frac{r'n_{\mathrm{e,i}}(r',a'+ct_{\mathrm{sf}}(a))}{\sqrt{r^2+r'^2-2rr'\cos\varphi+(a-a')^2}}.
\end{equation}
This formula characterizes the electric potential at any \textit{fixed} time during the development of the shower. Choosing the time at which the depth of the shower front is around $X_\mathrm{max}$, the electric field derived from $\mathbf{E}=-\mathbf{\nabla}V$ is the maximum one that is generated during the development of an air shower. Within a distance of $\simeq 100~$m from the shower axis, this leads to similar values for both the radial and the longitudinal components, namely of the order of 1~V~m$^{-1}$ for a shower with primary energy of $10^{17.5}~$eV. This value increases almost linearly with the primary energy. Hence, the approximations adopted so far are justified for showers induced by cosmic rays with energies up to 
$10^{19.5}~$eV. On the flip side, at higher energies, the decomposition of the global solution to equation~(\ref{eqn:boltzmann_general}) in terms of $f_{\mathbf{V}}^{(0)}+f_{\mathbf{V}}^{(1)}$ is likely not rigorous in some space regions around $X_\mathrm{max}$ and close to the shower axis.

\subsection{Plasma Absorption Effects}
\label{dielectric-shower}

To determine the absorption length vector in the plasma, we consider the electromagnetic interaction neglected so far between the radiation field and the electrons in the Boltzmann equation. In the same way as in equation~(\ref{eqn:boltzmann_perturb}), the magnetic term can be neglected since the amplitude of the magnetic field is smaller than the one of the electric field by a factor $c$ for an electromagnetic wave. The perturbation to the velocity distribution function caused by the radiation is thus similar to the solution previously obtained:
\begin{equation}
\label{eqn:solrad_perturb}
f^{(\mathrm{rad})}_\mathbf{V}(\mathbf{v},t)=-\frac{e}{m}\int_0^t~\dif t'~\left(\mathbf{E}^{\mathrm{rad}}(t')\cdot\mathbf{\nabla}_\mathbf{v} f_\mathbf{V}^{(0)}(v,t')\right)\exp{\left(\xi(v,t')-\xi(v,t)\right)}.
\end{equation}
The current vector $\mathbf{j}$ is then deduced from $\mathbf{j}(t)=e\int\dif \mathbf{v}~n_{\mathrm{e,i}}(v,t)\mathbf{v}f^{(\mathrm{rad})}_\mathbf{V}(\mathbf{v},t)$. To deduce an approximate expression of the conductivity tensor $\underline{\sigma}(\omega)$, we replace the time-dependent vector $\mathbf{\nabla}_\mathbf{v} f_\mathbf{V}^{(0)}(v,t')$ in the integrand of equation~(\ref{eqn:solrad_perturb}) by its time-average value and the argument of the exponential by $-\overline{\nu}_{\mathrm{eff}}(v)(t-t')$, with  $\overline{\nu}_{\mathrm{eff}}(v)$ the time-average value of the  $\nu_{\mathrm{eff}}(v,t)$ function. Besides, the density $n_{\mathrm{e,i}}$ is taken at $t=0$, where it is maximal. In this way, $\mathbf{j}$ appears as a convolution product and the Fourier transform of $\mathbf{j}$ is related to the Fourier transform of $\mathbf{E}^{\mathrm{rad}}$ through the expression:
\begin{equation}
\label{eqn:j_shower}
\tilde{\mathbf{j}}(\omega)=\frac{e^2}{m}\int\dif \mathbf{v}~n_{\mathrm{e,i}}^0(v)\mathbf{v}~\frac{\overline{\nabla_{\mathbf{v}}f_\mathbf{V}^{(0)}}(v)\cdot\tilde{\mathbf{E}}^{\mathrm{rad}}(\omega)}{i\omega+\overline{\nu}_{\mathrm{eff}}(v)}.
\end{equation}
This leads to the following dielectric coefficient\footnote{From equation~(\ref{eqn:j_shower}), the conductivity tensor $\underline{\sigma}$ is diagonal and proportional to \underline{1}; and so is the dielectric tensor.}:
\begin{equation}
\label{eqn:j_epsilon}
\epsilon(\omega)=1+\frac{4\pi e^2}{3m\epsilon_0}\int\dif v~v^4\frac{n_{\mathrm{e,i}}^0(v)}{\overline{\nu}_{\mathrm{eff}}^2(v)+\omega^2}\frac{\overline{\partial f_\mathbf{V}^{(0)}}}{\partial v}+i\frac{4\pi e^2}{3m\omega\epsilon_0}\int\dif v~v^4\frac{n_{\mathrm{e,i}}^0(v)\overline{\nu}_{\mathrm{eff}}(v)}{\overline{\nu}_{\mathrm{eff}}^2(v)+\omega^2}\frac{\overline{\partial f_\mathbf{V}^{(0)}}}{\partial v}.
\end{equation}

Applied to the reference shower, this leads to an absorption length of the order of $10^{8}~$m for densities close to the shower axis and to the depth of maximum of the showers. For a primary energy of $10^{21}~$eV, $L$ is never smaller than $10^{5}~$m close to the depth of maximum and close to the axis. Hence, plasma absorption effects can be ignored in terms of photon propagation from their production points to the ground at gigahertz frequencies in extensive air showers. Consequently, the attenuation factor $\exp{(-\kappa)}$ is set to 1 in the following.

\subsection{Application to the Reference Shower}
\label{application}

\begin{figure}[!h]
\centering
\includegraphics[width=12cm]{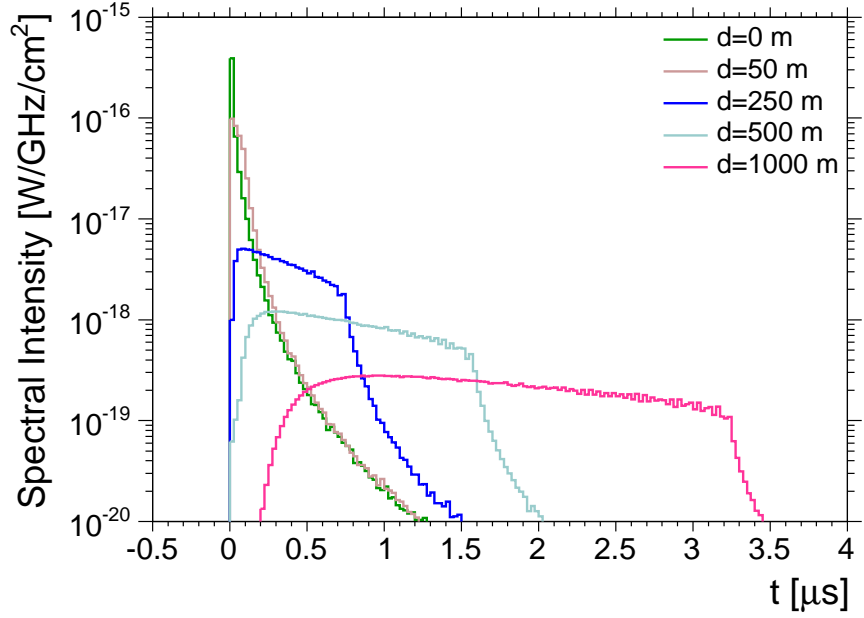}
\caption{\small{Spectral intensity $\Phi_{\mathrm{g}}^{\mathrm{e,i}}$ as a function of time expected at different distances from the shower core at ground level, for a vertical shower with energy $10^{17.5}~$eV.}}
\label{fig:sintensity}
\end{figure}

To probe the spectral intensity expected in the case of the vertical proton shower with primary energy of $10^{17.5}~$eV, we calculate equations~\ref{eqn:spectral_intensity_1} and~\ref{eqn:spectral_intensity_2} by taking the widely-used parameterization of the atmosphere \textit{US standard atmosphere}, based on experimental data~\cite{NASA-atm}. The parameterization of the wavelength dependence of the refractive index is taken from~\cite{Weast}.

The spectral intensity expected at different distances from the shower core at the sea level and averaged in time bins of 25~ns is shown in figure~\ref{fig:sintensity}. Amplitudes are rapidly decreasing for increasing distances to the shower core. Close to the core, the duration of the signals is controlled by the behaviour of the $f$ function. As the distance to the core increases, the duration of the signals is then controlled by the geometry of the shower through the different regions that can be probed for different positions of observers. The spectral intensity at ten kilometers from the core is often used as a reference value due to the scaling law put forward in~\cite{Gorham2008} to 'convert' laboratory measurements to extensive air shower expectations. We find here that $\Phi_{\mathrm{g}}^{\mathrm{e,i}}\simeq 2\times10^{-21}~$W~cm$^{-2}$~GHz$^{-1}$, which is about two times lower than the previous estimate in~\cite{AlSamarai2015} due to the new effects entering in the calculation of the $f$ function, and two orders of magnitude lower than the estimate presented in~\cite{Gorham2008}. 

A relevant estimate of the minimal spectral intensity $\Phi_{\mathrm{min}}$ detectable by an antenna operating in a bandwidth $\Delta\nu$ with a noise temperature $T_{\mathrm{sys}}$ and an effective area $A_{\mathrm{eff}}$ is known to obey
\begin{equation}
\Phi_{\mathrm{min}}=\frac{kT_{\mathrm{sys}}}{A_{\mathrm{eff}}\sqrt{\tau\Delta\nu}},
\end{equation}
where $k$ is the Boltzmann constant and $\tau$ the receiver sampling time. For values $\Delta\nu=0.8~$GHz, $\tau=10~$ns, and $T_{\mathrm{sys}}=50~$K, values typical of the current setups used at the Pierre Auger Observatory for instance~\cite{Gaior2013}, our results imply that the signals of the proxy shower can be detected close to the core on the condition that $A_{\mathrm{eff}}$ is of the order of $10^3$~cm$^2$. Far from the core, much larger collection areas are necessary. For larger shower energies, $A_{\mathrm{eff}}$ roughly decreases as the inverse of the primary energy of the cosmic ray.

\begin{figure}[!h]
\centering
\includegraphics[width=8.5cm]{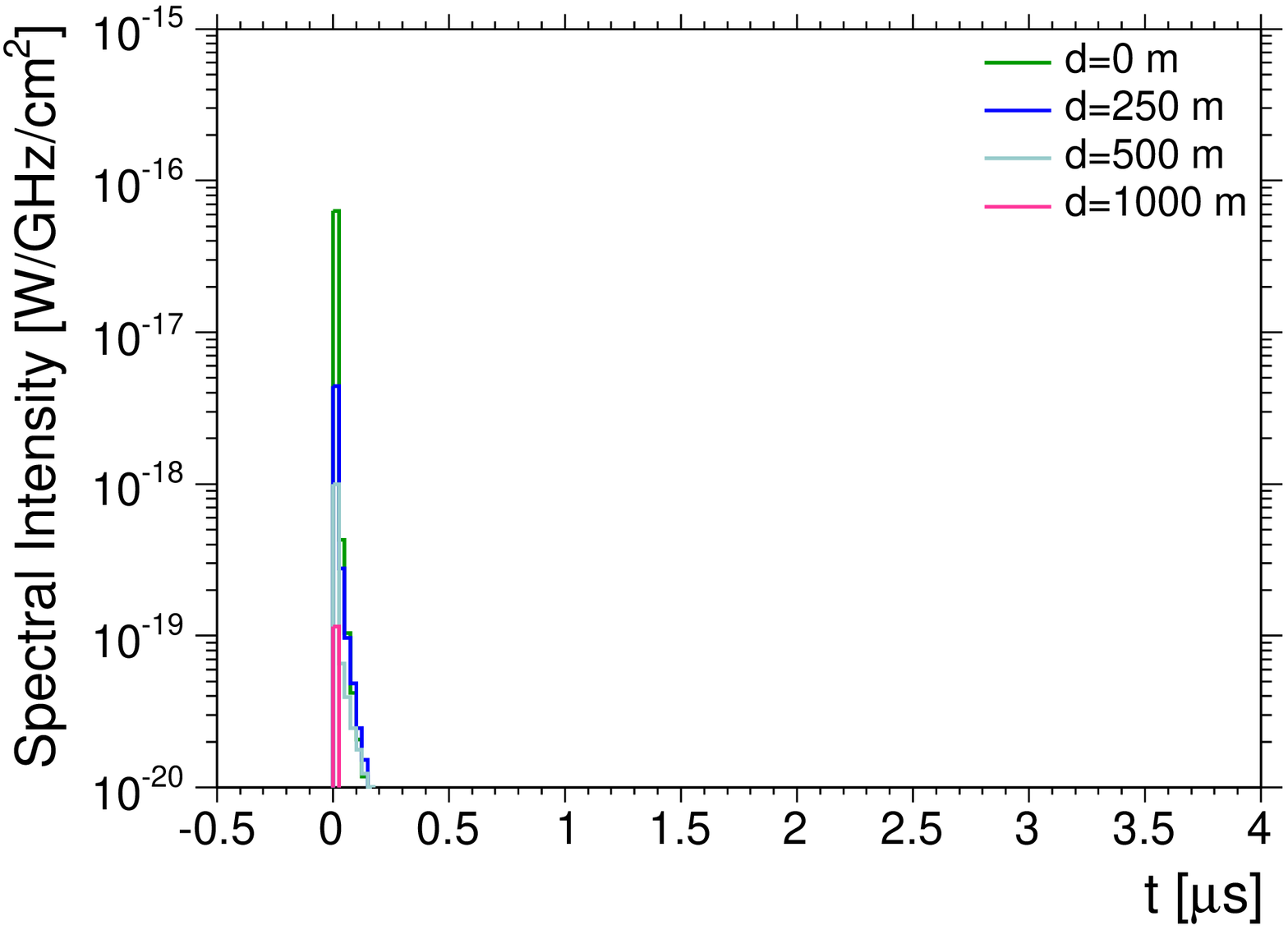}
\includegraphics[width=8.5cm]{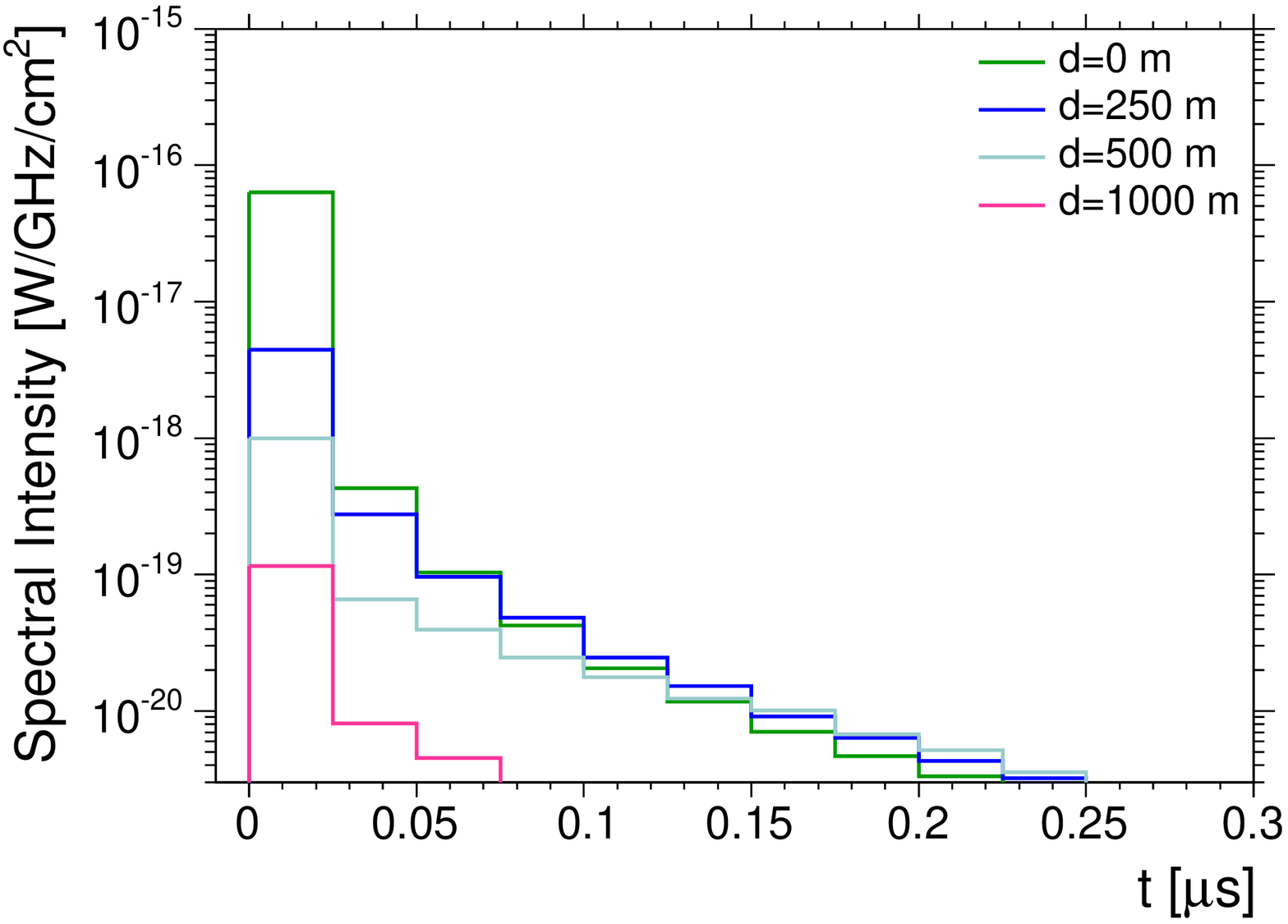}
\caption{\small{Spectral intensity $\Phi_{\mathrm{g}}^{\mathrm{e}}$ as a function of time expected at different distances from the shower core at ground level, for a vertical shower with energy $10^{17.5}~$eV. Left: same time scale as in figure~\ref{fig:sintensity}, for comparison purpose. Right: zoom of the distributions in time.}}
\label{fig:sintensity_prim}
\end{figure}

The spectral intensity $\Phi_{\mathrm{g}}^{\mathrm{e}}$ expected from the high-energy electrons of the shower front is shown in figure~\ref{fig:sintensity_prim}. Even though there are many fewer high-energy electrons than low-energy ones, it turns out that $\Phi_{\mathrm{g}}^{\mathrm{e}}$ reaches the same order of magnitude as $\Phi_{\mathrm{g}}^{\mathrm{e,i}}$ in the first nanosecond bins. This is due to the strong anisotropy in the Bremsstrahlung cross section for high-energy electrons, which concentrates the production of photons in the forward direction of the electrons.

\section{Molecular Bremsstrahlung Radiation in Air from a 95~keV Electron Beam}
\label{mbr2}

\subsection{The Experiment}
\label{experiment}

Recently, measurements of the microwave emission from a 95~keV electron beam in air in the Ku bandwidth ($\simeq$ 11~GHz) have been performed~\cite{Conti2014}. The power signal detected in the forward direction at $\simeq 30~$cm of the emission zone amounts to $\simeq 10^{-15}~$W for a beam current of $I=50~\mu$A. In this section, we adapt the formalism developed in previous sections to the specific setup of this experiment to check whether the order of magnitude of the emission can be reproduced. 

In this experiment, an electrostatic gun accelerates electrons up to 95~keV in pulses 150~$\mu$s long. Electrons exit the gun in air through a synthetic diamond window, 20~$\mu$m thick. In air, the electron energy spectrum follows a Landau distribution with most probable energy of 81~keV. The emitted signal is detected by a Low Noise Block (LNB) operating from 10.95~GHz to 11.70~GHz. The LNB feed is a pyramidal horn with 20~dB nominal gain and $\simeq 60~$\% aperture efficiency, positioned inside an anechoic chamber. The LNB has a single dipole antenna and detects one polarization each time. The dependence of the signal upon the beam current is compatible within uncertainties for the two polarizations, showing that the measured radiation is not polarized as expected in the case of Bremsstrahlung emission. In addition, successful tests showing that the signal is unambiguously generated by the electrons in air were pursued and are detailed in~\cite{Conti2014}. The signal can thus be unambiguously attributed, for the first time in a background free, controlled, and repeatable environment, to molecular Bremsstrahlung radiation. 

\subsection{Monte-Carlo Simulation}
\label{mc}

\begin{figure}[!t]
\centering
\includegraphics[width=15cm]{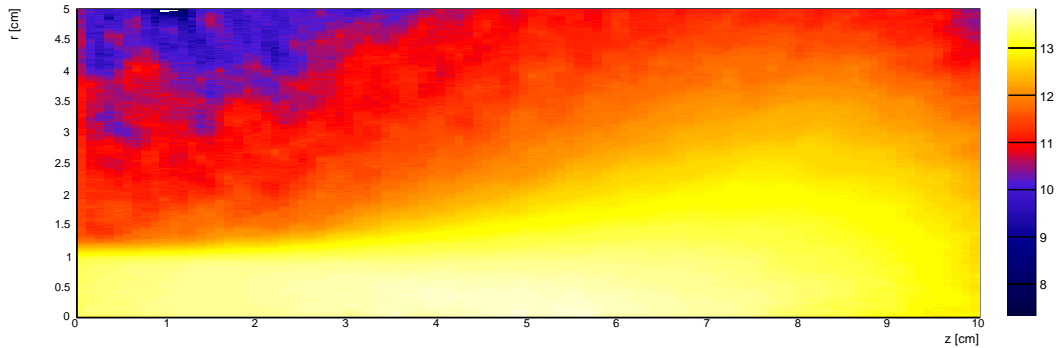}
\caption{\small{Average number of ionization electrons (in decimal logarithmic scale and inverse cubic meter unit) for a beam current of 50 $\mu$A, as a function of the position in air at the exit of the gun.}}
\label{fig:ne_conti}
\end{figure}

To probe the longitudinal and lateral extensions of the electrons left in air during each pulse, we defer to a Monte-Carlo simulation to estimate  the electron density as a function of the position in space. For this purpose, a large number of electrons is injected in air within a disk of radius\footnote{This radius is intended to account roughly for the spot size of the injected electrons by the gun and the spreading effect of the diamond window.} 1~cm. The energy of each electron is sampled from a truncated Landau distribution such that the most probable value is 81~keV. Electrons are then followed in elementary time steps $\delta t$ much smaller than the characteristic time scales of each interaction process of interest, namely ionization, excitation, attachment, and elastic collisions. Electrons are deflected mainly through elastic collisions. The corresponding angular differential cross section is taken from parameterizations from~\cite{Arqueros}. The whole cascade of secondary electrons created through ionization processes is simulated by following low-energy electrons one by one, until their disappearance through the attachment process. 

The average electron density $n_{\mathrm{e}}^1(\mathbf{v},\mathbf{r},t)$ per velocity band for one incoming electron exiting the gun is obtained from this simulation. Once the system is in its stationary state, the electron density $n_{\mathrm{e}}(r,z)$ at any time is obtained by the convolution of $n_{\mathrm{e}}^1(\mathbf{v},\mathbf{r},t)$ with the rate of incoming electrons from the gun and by marginalization over the velocities:
\begin{equation}
n_{\mathrm{e}}(r,z)=\int \dif \mathbf{v}\dif t'~n_{\mathrm{e}}^1(\mathbf{v},r,z,t')~\frac{I(t-t')}{e}=\frac{I}{e}\int \dif \mathbf{v}\dif t'~n_{\mathrm{e}}^1(\mathbf{v},r,z,t').
\end{equation}
This leads to densities as high as $\simeq 10^{14}~$m$^{-3}$ within a cylinder of a few centimeters of longitudinal extension and one centimeter of radial extension, as seen in figure~\ref{fig:ne_conti}. The electric field generated by such a density is much larger than the one considered in the previous section in the case of extensive air showers. To simplify the discussion, we consider only the radial component of this field in the following, which, given the geometry, is expected to dominate in most of the volume:
\begin{equation}
\label{eqn:efield_conti}
\mathbf{E}(r,z)=\frac{e}{r\epsilon_0}\int_0^r\dif r'~r'n_{\mathrm{e}}(r',z)~\mathbf{u}_r.
\end{equation}
This yields a field amplitude of about $10^3$~V~m$^{-1}$ a few centimeters away from the axis of symmetry.

\subsection{Bremsstrahlung Radiation}
\label{mbr-conti}

The large values of the electric field, in contrast to the case of extensive air showers, prevents us from using the same solution for $f_{\mathbf{V}}$ as the one of the previous section. It is worth noting that the use of the same solution would lead to an expected power of the order of $10^{-13}~$W, that is, two orders of magnitude above the observed value. 

So, given the large values of the electric field, the electromagnetic term is here dominant. To solve the Boltzmann equation including both the electromagnetic and the collision terms, we consider a simplified collision term\footnote{Note that by using the simplified effective collision rate $\overline{\nu}_\mathrm{eff}(v)$ in the absence of the electromagnetic term, the expected power is smaller than the one obtained with the complete effective collision rate $\nu_{\mathrm{eff}}(v,t)$ by a factor $\simeq 2.5$. This gives an order of magnitude of the error induced by this simplification when considering the electromagnetic term, although this error factor may be affected by this additional term.} obtained by replacing $\nu_{\mathrm{eff}}(v,t)$ by its time-average value $\overline{\nu}_\mathrm{eff}(v)$:
\begin{equation}
\label{eqn:boltzmann_conti}
\frac{\partial f_\mathbf{V}(\mathbf{v},\mathbf{r},t)}{\partial t}+\frac{eE_r(\mathbf{r})}{m}\frac{\partial f_\mathbf{V}(\mathbf{v},\mathbf{r},t)}{\partial v_r}=-\overline{\nu}_\mathrm{eff}(v) f_\mathbf{V}(\mathbf{v},\mathbf{r},t).
\end{equation}
Although an explicit spatial dependence of the distribution $f_{\mathbf{V}}$ is introduced due to the corresponding dependence of the electric field, the spatial dispersion term is still small and neglected. We note, moreover, that the expression~(\ref{eqn:efield_conti}) for the electric field relies on the one of the electron density $n_{\mathrm{e}}(\mathbf{r})$ and thus of $n_{\mathrm{e}}^1(\mathbf{v},\mathbf{r},t)$, while the comprehensive determination of $n_{\mathrm{e}}^1(\mathbf{v},\mathbf{r},t)$ should account for the electric field to be determined. This would lead to a complex system of coupled equations. The strategy adopted here, consisting of neglecting $\mathbf{E}$ to determine $n_{\mathrm{e}}^1$, can be viewed as the first step of an iteration procedure. 

The solution for the Laplace transform $X(p,\mathbf{r},\mathbf{v})$ of equation~(\ref{eqn:boltzmann_conti}) yields to the following expression:
\begin{equation}
X(p,\mathbf{r},\mathbf{v})=\frac{m}{4\pi eE_r}\int_0^{v_{\mathrm{max}}}\dif v_r'~\frac{f_0(\sqrt{v_r'^2+v_z^2})}{v_r'^2+v_z^2}\exp{\left(-\frac{m}{eE_r(\mathbf{r})}\int_{v_r'}^{v_r}\dif v_r''~\left(p+\overline{\nu}_\mathrm{eff}\left(\sqrt{v_r''^2+v_z^2}\right)\right)\right)}.
\end{equation}
The function $f_\mathbf{V}(\mathbf{v},t)$ is then obtained by inverse Laplace transform, using the Stehfest algorithm~\cite{Stehfest1970}. 
From the knowledge of $f_\mathbf{V}(\mathbf{v},t)$, the spectral intensity for one incident electron from the gun, $\Phi_1(t)$, is estimated in the same way as in the previous section, by adapting the formalism to the specific geometry of this experiment. Here, the instantaneous density of ionization electrons left after the passage of one incoming electron from the gun is denoted by $n_{\mathrm{e}}^{1,0}(\mathbf{r},t)$ and is extracted from the Monte-Carlo simulation. This leads to the following expression:
\begin{equation}
\label{eqn:phi1-conti}
\Phi_1(t)=\frac{h\rho\mathcal{N}\pi^2}{4A}\int_0^{r_\mathrm{max}}\int_0^{z_\mathrm{max}}\int_0^{v_{\mathrm{max}}}\int_{-\sqrt{v^2_\mathrm{max}-v_r^2}}^{\sqrt{v^2_\mathrm{max}-v_r^2}}\dif r\dif z\dif v_r\dif v_z~ rv_r\frac{\sqrt{v_r^2+v_z^2}}{R^2(r,z)}n_{\mathrm{e}}^{1,0}(\mathbf{r})f_{\mathbf{V}}(\mathbf{v},\mathbf{r},t_{\dif }(t))\Theta(t_{\dif }(t))\frac{\nu\dif \sigma}{\dif \nu},
\end{equation}
with $R$ the distance between the antenna and the emission point, and $t_{\dif }(t)=t-Rn/c-t_0(\mathbf{r})$ is the delayed time at which the emission started after the passage of the incident electron of the gun at time  $t_0(\mathbf{r})$. Once the system is in its stationary state, the expected power $P$ at any time is obtained by the convolution of $\Phi_1(t)$ with the rate of incoming electrons and by coupling the resulting spectral intensity to the effective area $\lambda^2/4\pi$ and bandwidth $\Delta\nu$ of the antenna of gain pattern $G$:
\begin{equation}
\label{eqn:power_conti}
P=\frac{1}{2}\frac{G\Delta\nu\lambda^2}{2\pi}\int \dif t'~\Phi_1(t')~\frac{I(t-t')}{e}=\frac{1}{2}\frac{G\Delta\nu\lambda^2I}{2\pi e}\int \dif t'~\Phi_1(t'),
\end{equation}
where the factor $1/2$ accounts for the random polarization of the Bremsstrahlung radiation (polarization loss). Given the geometry, the gain factor $G$ is taken in the forward direction only. 

\begin{figure}[!h]
\centering
\includegraphics[width=12cm]{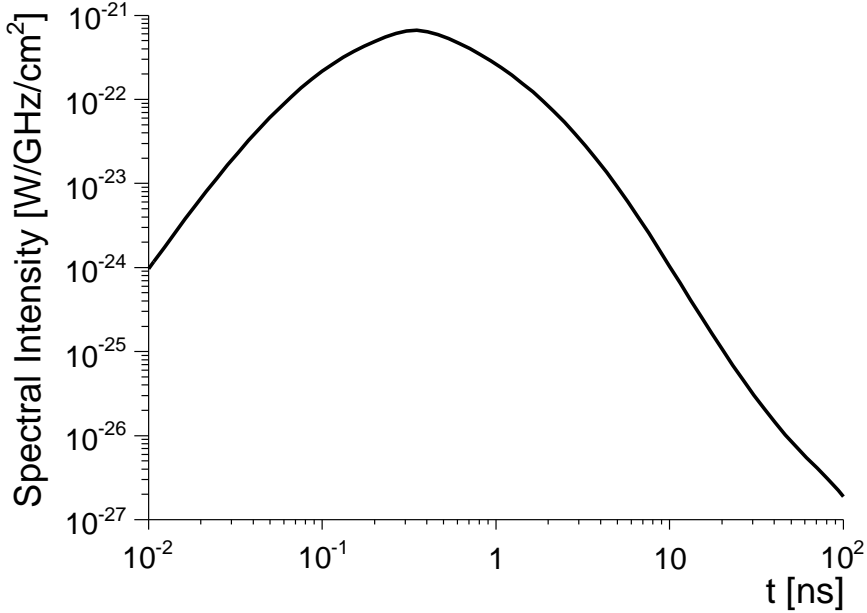}
\caption{\small{Average spectral intensity expected at 30~cm from the beam region for one incident electron emitted from the gun.}}
\label{fig:phi_conti}
\end{figure}

The spectral intensity $\Phi_1(t)$ is shown in figure~\ref{fig:phi_conti}. It is observed to decrease after less than 1~ns, which is a time scale smaller than the one expected from collisions. This is due to the influence of the electromagnetic interactions. In a simple picture, due to their mutual electric repulsion, electrons are decelerated when they approach each other so that in average, they spend longer times with an energy in the range where they get attached at a rate larger than 1~GHz than in the regime where the electromagnetic interactions are negligible.

Once $\Phi_1(t)$ is plugged into equation~(\ref{eqn:power_conti}), the resulting expected power is $\simeq 2.5\times10^{-15}~$W, which is close to the observed value. While the agreement is not perfect between this estimate and the measurement, it is worth noting that the order of magnitude is reproduced. The various approximations performed to get at this estimate could be tackled through an iterative procedure, but this would require an exact description of the experimental conditions which turns out to be very complex. Hence, we restrict ourselves to this estimate. 

It is also interesting to look at the expected power as a function of the beam current $I$. In~\cite{Conti2014}, the increase of the power with the beam intensity is observed to be linear but not strictly proportional, the value for $I=250~\mu$A being $\simeq 4.4~$ larger than for $I=50~\mu$A. In the framework developed in this study, the power is indeed not expected to be strictly proportional to the beam current (and could even be non linear in a regime of high current). Repeating the calculations for $I=250~\mu$A, the expected power is found to be 4 times larger than the one found for $I=50~\mu$A, which also agrees well with the observations.

\subsection{Additional Radiation?}
\label{comments}

In addition to the Bremsstrahlung emission, electrons radiate as the result of their acceleration/deceleration in the surrounding electric field they generate. The intensity of the emission from one electron with velocity $\mathbf{v}$ received at any distance $R$ obeys the Larmor formula. By making use of the Parseval identity, the frequency spectrum of the radiated energy per surface unit reads as
\begin{equation}
\frac{\dif E}{R^2\dif \nu}=\frac{e^2\cos^2{\theta}}{16\pi^2R^2\epsilon_0c^3}\left|\int_0^{\Delta t}\dif t~\dot{\mathbf{v}}(t)\exp{(-i2\pi\nu t)}\right|^2,
\end{equation}
with $\theta$ defined such that $\theta=0$ in the longitudinal direction and $\theta=\pi/2$ in the radial one. Here, the time integration interval $\Delta t$ corresponds to the time between two collisions (including elastic collisions): $\Delta t=\nu_\mathrm{c}^{-1}(v)$. Since the collision frequency is much larger than the frequency range analyzed here, the argument of the exponential remains small during the integration time scale so that the exponential can be approximated by 1. Neglecting also the change of position from $\mathbf{r}(0)$ to $\mathbf{r}(\Delta t)$ during $\Delta t$, the energy spectrum per surface unit can be approximated by
\begin{equation}
\frac{\dif E}{R^2\dif \nu}\simeq\frac{e^4\cos^2{\theta}}{16\pi^2R^2\epsilon_0c^3m^2\nu_\mathrm{c}^2(v)}\left|\mathbf{E}(\mathbf{r})\right|^2.
\end{equation}
The spectral intensity expected from this radiation is then obtained by coupling this expression to the collision rate per volume unit and per electron velocity band, and by integrating the resulting expression over $\mathbf{v}$ and $\mathbf{r}$:
\begin{equation}
\label{eqn:phi-rad}
\Phi_1^{\mathrm{rad}}(t)\simeq\frac{\rho\mathcal{N}e^4\cos^2{\theta}}{64\pi^3A\epsilon_0c^3m^2} \int\dif \mathbf{v}\int\dif \mathbf{r}~ \frac{v}{\nu_\mathrm{c}^2(v)R^2(r,z)}\sigma^{\mathrm{tot}}(v)n_{\mathrm{e}}^{1,0}(\mathbf{r})\left|\mathbf{E}(\mathbf{r})\right|^2f_{\mathbf{V}}(\mathbf{v},\mathbf{r},t_{\dif }(t))\Theta(t_{\dif }(t)).
\end{equation}
This spectral intensity turns out to be much smaller than equation~(\ref{eqn:phi1-conti}) for $\Phi_1(t)$ by several orders of magnitude. 

However, equation~(\ref{eqn:phi-rad}) has been obtained under the underlying assumption of an incoherent emission. It is worth estimating the same quantity in the case of a coherent one, since the electromagnetic fields of the radiated photons produced at the same time can add up coherently for electrons separated by a distance less than the considered wavelength. In this case, the spectral intensity can be reasonably estimated as:
\begin{equation}
\label{eqn:phi-rad-coh}
\Phi_{1,\mathrm{coh}}^{\mathrm{rad}}(t)\simeq\frac{\rho\mathcal{N}e^4\cos^2{\theta}}{128\pi^3A\epsilon_0c^3m^2} \int\dif \mathbf{v}\int\dif \mathbf{r}\int\dif \mathbf{r^\prime}~ \frac{v}{\nu_\mathrm{c}^2(v)R^2(r,z)}\sigma^{\mathrm{tot}}(v)n_{\mathrm{e}}^{1,0}(\mathbf{r})n_{\mathrm{e}}(\mathbf{r^\prime})\left|\mathbf{E}(\mathbf{r})\right|^2f_{\mathbf{V}}(\mathbf{v},\mathbf{r},t_{\dif }(t))\Theta(t_{\dif }(t)),
\end{equation}
where the integration over $\mathbf{r^\prime}$ is limited to small volumes around the current integration position $\mathbf{r}$, volumes of the order of the cube of the considered wavelength. Equation~(\ref{eqn:phi-rad-coh}) turns out to lead to a spectral intensity whose maximum is of the same order of magnitude as $\Phi_1(t)$, and consequently to a power $P$ of the order of $1.5\times10^{-15}~$W. Although we do not claim to be able to reproduce the exact experimental conditions, in particular concerning the exact geometry and density of electrons in air, it is interesting that this rough estimate shows that part of the observed power could be attributed to this radiation in addition to the Bremsstrahlung one. Interestingly, the angular dependence of the radiation studied here, in $\cos^2{\theta}$, would fit remarkably well the observed one.

\subsection{Absorptions Effects}
\label{comments2}

Eventual absorption effects have been neglected to get at equations~\ref{eqn:phi-rad} and~\ref{eqn:phi-rad-coh}. In this case, the approach followed in section~\ref{dielectric-shower} can only be reproduced numerically and thus prevents us from getting explicit expressions for the dielectric tensor. However, an approximated expression can be obtained by approaching $\left\langle \mathbf{v}\right\rangle$ with the simplified equation of motion for the electron momentum $\mathbf{p}$:
\begin{equation}
\frac{\mathrm{d^2}\mathbf{p}}{\dif t^2}=-\nu_{\mathrm{c}}\frac{\dif \mathbf{p}}{\dif t}-\omega_p^2\left(\mathbf{p}\cdot\mathbf{u}_r\right) \mathbf{u}_r-ie\omega\mathbf{E},
\end{equation}
where the effect of the electric field is simplified in terms of an elastic restoring force with pulsation $\omega_p=\sqrt{n_{\mathrm{e}}^{1,0}e/m\epsilon_0}$ (the plasma pulsation).  The steady-state solution for $\left\langle\mathbf{p}\right\rangle$ leads to the following expression for the dielectric tensor:
\begin{eqnarray}
\label{eqn:eps-plasma}
\left[\epsilon(\omega,\mathbf{r})\right]_{xx}=\left[\epsilon(\omega,\mathbf{r})\right]_{yy}=1-\frac{e^2}{m\epsilon_0\omega^2} \int \dif T~\frac{n_{\mathrm{e}}(T,\mathbf{r})(\omega/\nu_{\mathrm{c}}(T))^2}{1+(\omega/\nu_{\mathrm{c}}(T))^2}+i~\frac{e^2}{m\epsilon_0\omega^2}\int \dif T~\frac{n_{\mathrm{e}}(T,\mathbf{r})(\omega/\nu_{\mathrm{c}}(T))}{1+(\omega/\nu_{\mathrm{c}}(T))^2}, \nonumber \\
\left[\epsilon(\omega,\mathbf{r})\right]_{zz}=1+\frac{e^2}{m\epsilon_0\omega\tilde{\omega}} \int \dif T~\frac{n_{\mathrm{e}}(T,\mathbf{r})(\tilde{\omega}/\nu_{\mathrm{c}}(T))^2}{1+(\tilde{\omega}/\nu_{\mathrm{c}}(T))^2}+i~\frac{e^2}{m\epsilon_0\omega\tilde{\omega}}\int \dif T~\frac{n_{\mathrm{e}}(T,\mathbf{r})(\tilde{\omega}/\nu_{\mathrm{c}}(T))}{1+(\tilde{\omega}/\nu_{\mathrm{c}}(T))^2},
\end{eqnarray}
with in shortened notation $\tilde{\omega}=\omega_p^2/\omega-\omega$. At 10~GHz, minimal values of the radial component of the absorption length vector $\mathbf{L}(\nu,\mathbf{r})$ are found in the core of the beam, where the density of electrons is the largest. These values are of the order of hundreds of meters, much longer than the radial and longitudinal sizes of the beam by orders of magnitude. The longitudinal component of $\mathbf{L}(\nu,\mathbf{r})$ is very similar. Hence, plasma dispersion effects have no impact in this experiment.

\section{Conclusion}
\label{conclusion}

In this paper, we have presented a complete model to estimate the spectral intensity expected at ground level from molecular Bremsstrahlung radiation in extensive air showers. To this aim, a Boltzmann equation has been used to determine the time evolution of the ionization electrons produced after the passage of the cascade in the atmosphere. We have shown that the collision term is the dominant one in this equation, and have performed an exhaustive treatment of the different processes entering into this collision term. The absorption of the emitted photons in the plasma have been shown to be negligible. In addition, we have also estimated the spectral intensity expected from the molecular Bremsstrahlung radiation of the high-energy electrons of the shower. It turns out that, close to the shower core, both sources of radiation provide the same amount of expected signal during the first nanoseconds. The spectral intensity at ten kilometers from the core, which is often used as a reference value, is found to be $\Phi_{\mathrm{g}}\simeq 2\times10^{-21}~$W~cm$^{-2}$~GHz$^{-1}$. This is about two times lower than the previous estimate in~\cite{AlSamarai2015} due to the new effects entering in this study. From an experimental point of view, such a value requires significant enhancements of sensitivity to be detected.

We have also applied our model to the case of the recent laboratory setup described in~\cite{Conti2014}, where molecular Bremsstrahlung radiation from electrons in air has been unambiguously measured. We have shown that the density of electrons, much higher in this experiment than in extensive air showers, induces effects that require consideration of the electromagnetic interactions of the electrons on top of their collisions. The net effect is, in average, to reduce the duration of the emission for each electron. The measured value can then be reasonably reproduced by the model. This underlines, however, \textit{the difficulty to extrapolate laboratory measurements to expectations for extensive air showers}, where only collisions are expected to impact the kinetic energy distribution of the electrons. 

Experimental setups using regularly spaced antennas oriented vertically, or nearly vertically, to detect showers crossing the field of view of the receivers have been deployed over the past years: the EASIER installation at the Pierre Auger observatory~\cite{Gaior2013} and at the CROME experiment~\cite{CROME}. The sensitivity of these installations is unfortunately below or at the edge of the one required to detect the molecular Bremsstrahlung radiation presented in this study. However, a few events have been interestingly detected in the past years through the observation of very short signals in time for antennas close to the shower core. These events could be due to geomagnetic and Askaryan mechanisms responsible for radio-emission in extensive air showers in the megahertz frequency range, mechanisms which have been shown to extend their emission in the gigahertz range in an elliptical ring-like region around the intersection of a Che\-ren\-kov cone with its vertex at shower maximum and the ground. This is due to the coherence of the emission occurring with relative time delays with respect to the virtual front plane moving at the speed $c$ within a few nanoseconds inside a thin cylinder around the shower maximum~\cite{AlvarezMuniz2011,Werner2012,AlvarezMuniz2012}. A correlation between the observed polarization of the signals and the polarization signatures typical of the geomagnetic/Askaryan emissions has been found at the CROME experiment~\cite{CROME}, providing some support that the geomagnetic/Askaryan emission mechanisms are responsible for an important fraction of the detected signals. Meanwhile, we note that the observed East/West asymmetry of the observed signal strength at the CROME experiment does not allow one to favor the extension of geomagnetic and Askaryan mechanisms in the gigahertz range as the emission mechanism over the molecular Bremsstrahlung radiation. This is because this asymmetry is a consequence of the combination of the excess of electrons over positrons in the shower front and of the magnetic deflections in opposite directions for electrons and positrons (effects which have not been considered in this study). Consequently, any radiation from the electrons/positrons is expected to show the same kind of asymmetry. Hence, the effects studied in this paper could participate to some extent to the detection of these events, or could even explain the majority of these events. 

New installations with antennas pointing nearly vertically are currently running at the Pierre Auger observatory. The improved sensitivity of this installation due to higher gains and larger effective areas of the antennas should allow the collection of a much larger statistics in future years. This will help in understanding the sources of emission at gigahertz frequencies.

\section*{Acknowledgements}

We acknowledge the support of the French Agence Nationale de la Recherche (ANR) under reference ANR-12-BS05-0005-01. 



\end{document}